\documentclass[usenatbib,usegraphicx]{mn2e}

\catcode`\@=11
\newcommand\percubcm{\mbox{$\m@th\,\rm cm^{-3}$}}
\newcommand{\persecond}{\mbox{$\m@th\,\rm s^{-1}$}}
\newcommand{\cccs}{\mbox{$\m@th\,\rm cm^3\,s^{-1}$}}
\newcommand{\kms}{\mbox{$\m@th\,\rm km\,s^{-1}$}}
\newcommand{\K}{\mbox{$\m@th\,\rm K$}}

\newcommand{\alfaji}{\mbox{$\m@th\alpha^{\rm eff}_{ji}$}}
\newcommand{\alfaef}{\mbox{$\m@th\alpha^{\rm eff}_j$}}
\newcommand{\betaef}{\mbox{$\m@th\beta^{\rm eff}_{gj}$}}
\newcommand{\hb}{\mbox{$\m@th\rm H\beta$}}
\newcommand{\lam}{\mbox{$\m@th\lambda\,$}}
\newcommand{\lala}{\mbox{$\m@th\lambda\lambda\,$}}
\newcommand{\term}[1]{\mbox{$\m@th\rm #1$}}
\newcommand{\mult}[2]{$\m@th\rm #1$--$\m@th\rm #2$}

\newcommand{\hii}{\mbox{H\thinspace{\sc ii}}}
\newcommand{\hp}{\mbox{$\m@th\rm H^+$}}

\newcommand{\cii}{\mbox{C\thinspace{\sc ii}}}
\newcommand{\cp}{\mbox{$\m@th\rm C^+$}}
\newcommand{\cpp}{\mbox{$\m@th\rm C^{2+}$}}
\renewcommand{\ni}{\mbox{N\thinspace{\sc i}}}
\newcommand{\nii}{\mbox{N\thinspace{\sc ii}}}
\newcommand{\np}{\mbox{$\m@th\rm N^+$}}
\newcommand{\npp}{\mbox{$\m@th\rm N^{2+}$}}
\newcommand{\oi}{\mbox{O\thinspace{\sc i}}}
\newcommand{\oii}{\mbox{O\thinspace{\sc ii}}}
\newcommand{\oiii}{\mbox{O\thinspace{\sc iii}}}
\newcommand{\op}{\mbox{$\m@th\rm O^+$}}
\newcommand{\opp}{\mbox{$\m@th\rm O^{2+}$}}

\newcommand{\beq}{\begin{equation}}
\newcommand{\eeq}{\end{equation}}
\newcommand{\ecua}[1]{equation~(\ref{#1})}

\newcommand{\clo}{{\sc cloudy}}
\newcommand{\clotd}{{\sc cloudy\_3D}}
\newcommand{\cmf}{{\sc cmfgen}}
\newcommand{\emi}{{\sc emili}}
\newcommand{\ic}{{IC~418}}

\newcommand{\aap}{A\&A}
\newcommand{\aaps}{A\&A Suppl.}

\newcommand{\apj}{ApJ}
\newcommand{\apjl}{ApJ Letters}
\newcommand{\apjs}{ApJS}
\newcommand{\mnras}{MNRAS}
\newcommand{\rmaa}{RMAA}
\newcommand{\pasp}{PASP}

\catcode`\@=12

\title[Fluorescence and recombination in IC418]
{Excitation of emission lines by fluorescence 
and recombination in IC418}
\author[Escalante, Morisset and Georgiev]
{V. Escalante$^1$\thanks{E--mails: 
v.escalante@crya.unam.mx;\hfill\break
chris.morisset@gmail.com; 
georgiev@astro.unam.mx},
C. Morisset$^{2,3}$\footnotemark[1]
and L. Georgiev$^2$\footnotemark[1]\\
$^1$Centro de Radioastronom{\'\i}a y Astrof{\'\i}sica, 
Universidad Nacional Aut\'onoma de M\'exico, 
Ap.~Postal 72--3, C.~P.~58091, \\
Morelia, Michoac\'an, M\'exico\\ 
$^2$Instituto de Astronom{\'\i}a, 
Universidad Nacional Aut\'onoma de M\'exico, 
Ap.~Postal 70--264, C.~P.~04510, M\'exico, DF, M\'exico\\
$^3$Instituto de Astrof{\'\i}sica de Canarias, 
C/ V{\'\i}a L{\'a}ctea, S/N, 38205 San Crist{\'o}bal de la Laguna (Tenerife), 
Espa\~na}

\begin{document}

\pagerange{\pageref{firstpage}--\pageref{lastpage}}

\maketitle

\label{firstpage}

\begin{abstract}
We compare calculated intensities of lines of \cii, \ni, \nii\,  
\oi\ and \oii\ with a published deep spectroscopic 
survey of \ic. 
Our calculations use a self--consistent nebular model and 
a synthetic spectrum of the central star atmosphere to 
take into account line excitation by continuum 
fluorescence and electron recombination. 
We found that the \nii\ spectrum of s, p and most d states 
is excited by fluorescence 
due to the low excitation conditions of the nebula. 
Many \cii\ and \oii\ lines have significant amounts of 
excitation by fluorescence. 
Recombination excites all the lines from f and g states 
and most \oii\ lines. 
In the neutral--ionized boundary 
the \ni\ quartet and \oi\ triplet dipole allowed lines are 
excited by fluorescence, 
while the quintet \oi\ lines are excited by recombination. 
Electron excitation produces the forbidden optical lines of \oi\, 
and continuum fluorescence enhances the \ni\ forbidden line 
intensities. 
Lines excited by fluorescence of light below the Lyman limit 
thus suggest a new diagnostic 
to explore the inner boundary of the photodissociation region 
of the nebula. 
\end{abstract}


\section{Introduction}

With an apparent simple geometry 
the low excitation planetary nebula 
\ic\ lends itself to simpler models with fewer assumptions 
on its structure and kinematics 
than are generally 
necessary to reproduce observations of planetary nebulae (PNe). 
A recent deep spectroscopic survey of this object 
by \citet{emili} (hereafter SWBH) identified 
several hundred lines in \ic\ from which 
\citet*{sharpee} (hereafter SBW) 
derived ionic abundances for CNONe ions relative to H. 
Many dipole--allowed line intensities from those ions 
in that survey cannot be explained 
by the recombination theory with the same abundances derived 
from collisionally excited lines (CELs). 
The intensities of recombination lines are  
often used to measure ionic abundances 
because they depend more weakly on temperature 
and density than CELs. 
Abundances derived from recombination lines, however, 
are usually higher than those derived from CELs 
in the Orion nebula \citep*{esteban,baldwin00,esteban04,mesa},  
many \hii\ regions \citep{garcia07}, 
and planetary nebul\ae\ 
\citep*{liu95,liu01b,luo,liu06,tsamis08,garcia09,otsuka}. 
\citet{williams} showed that CELs probably give the 
correct abundances because the values derived from CELs 
are consistent with those derived from UV absorption lines in 
a sample of four PNe.  
The discrepancies 
of less than a factor of 2 
found by SBW 
in \ic\ between abundances from CELs and dipole--allowed lines 
are smaller than in other objects. 
Nevertheless it is important to ensure that 
the intensities of the dipole--allowed lines are not 
enhanced by additional excitation mechanisms besides 
recombination whenever they are used in temperature or 
density diagnostics \citep*{fang} (hereafter FSL) 
or abundance determinations. 

The usual procedure to determine a species X abundance 
directly from optical line intensities is to measure 
the ratio of the intensity of a line emitted by 
the species with respect to the \hb\ intensity 
and assume that the ratio of column densities 
producing the lines is equal to the abundance. 
Specifically, the abundance X/H is taken 
from the equation
\beq
{I(\lambda)\over I(\hb)}={\int\epsilon(\lambda)\,dV 
\over\int\epsilon(\hb)\,dV}=
{\langle\epsilon(\lambda)\rangle\over\langle\epsilon(\hb)\rangle}
{\rm X\over H}
\label{ionabun}
\eeq
where $I(\lambda)/I(\hb)$ is the ratio of intensities 
corrected for reddening, 
$\epsilon(\lambda)$ and $\epsilon(\hb)$ 
are the emissivities of a line emitted by X 
and of \hb\ respectively, 
both are integrated over some volume fraction of the nebula 
covered by the aperture, 
and $\langle\epsilon\rangle$ means some kind of average of the 
emissivities over that volume fraction. 
Averaged emissivities over the volume within the solid angle of 
the aperture imply the adoption of average values for the 
temperature, density and filling factors. 

An alternative procedure to determine chemical abundances 
in a nebula is to propose mutually consistent models of the 
nebula and the atmosphere of the exciting star or stars. 
The stellar parameters, chemical abundances and 
density profile of the nebula are inputs to a 
photoionization model, 
and their values are fixed by matching the 
predicted line intensities to the observations. 
The variation of stellar and nebular parameters of 
the model can be constrained by direct observations 
of the star and nebula to reduce the degeneracy of 
different parameters giving similar matches to the 
observed line intensities. 
\citet{chris09} (hereafter MG) showed that a consistent 
stellar energy distribution (SED) of the central star of \ic\ 
and a photoionization model give an impressive agreement of 
predictions and measured intensities of CELs and many 
recombination lines within observational errors. 
However most dipole--allowed lines in the SWBH survey 
were not included in the model by MG, 
and some of the ones that were included show 
observed intensities much larger than their predicted 
values because the MG calculations did not include 
fluorescence excitation. 

In this work we add fluorescence excitation of permitted 
lines to a photoionization model and a SED similar to the ones 
used by MG, 
and compare the results with 
all lines observed by SWBH that have probable identifications as 
dipole--allowed lines of \cii, \ni, \nii, \oi, and \oii\ and 
optical forbidden lines of \ni\ and \oi. 
Continuum fluorescence of starlight was suggested as a mechanism 
to excite permitted lines in ionized nebulae by \citet{seaton68} 
and \citet{grandi75,grandi76}. 
\citet{orion} (hereafter EM) showed that fluorescence in Orion 
dominates the intensities of \nii\ permitted lines. 
The low excitation level of \ic\  
and the resemblance of its ionization degree 
to that of the Orion nebula noted by \citet*{torres} 
suggest that a similar fluorescence mechanism may occur 
in both objects. 

\section{Line emissivities}

In our previous work (EM) we found that the rate of continuum
fluorescence excitation typically depends on the absorption of
photons in several resonant transitions, 
also called pumping transitions, 
to quantum states of moderate excitation.  
Excitation by recombination on the other hand involves the
capture of free electrons into a much larger number of excited states. 
In the fluorescence and recombination mechanisms subsequent transitions
to lower states produce the optical 
lines observed in ionized regions although
many lines in the UV and IR are also emitted.

Our calculation of quantum state populations is based on 
the cascade matrix formalism described in detail by EM. 
In that formalism excited states are populated 
by recombinations and by transitions from the ground 
and metastable states. 
The contribution of recombinations to 
the population of a state $j$ is given by the effective 
recombination coefficient of the state, $\alfaef$. 
The effective fluorescence coefficient defined by EM, 
$\betaef$, 
gives the contribution of pumping transitions from the 
ground and metastable states to excited states 
that produce decays ending in state $j$.
Both coefficients can be calculated in terms of the 
cascade matrix of \citet{seaton59}. 
The emissivity of a line for a transition from state $j$ to $i$ 
with a branching ratio $P_{ji}$ and frequency $\nu$ at 
a point in the nebula is
\beq
\epsilon_{ji}=h\nu P_{ji}(n_en^+\alfaef+\sum_gn_g\betaef)
\label{emis}
\eeq
where $n_e$ is the electron density, $n^+$ 
is the density of the ion before recombination, 
and $n_g$ is the population density of the ground 
or metastable state of the recombined ion or atom. 
The branching ratio $P_{ji}=A_{ji}/A_j$, 
where $A_{ji}$ is the Einstein coefficient for 
transition from states $j$ to $i$, 
and $A_j$ is the total decay rate of the upper state $j$. 
The {\it line} effective recombination coefficient is 
defined as: 
\beq
\alfaji = P_{ji}\alfaef
\label{alfaef}
\eeq
It is assumed that the ion density $n^+$ 
can be calculated from the ionization balance condition,  
and the ground and metastable state populations can be calculated 
from the balance of collisional and radiative transitions with 
the aid of a nebular photoionization model.

The cascade matrix formalism assumes that 
every transition from the continuum or from the 
ground or metastable state to a higher--energy state 
is followed by transitions to lower--energy states, 
and that metastable states are not significantly 
affected by those transitions. 
Those are valid assumptions at the temperatures 
and densities of a photoionized nebula because 
collisional excitation and deexcitation with 
electrons control the population of metastable 
states, 
but fluorescence can compete with 
collisions with electrons and atoms 
in the colder interface of the neutral and ionized 
part of a nebula.  
\citet{bautista} proposed that the forbidden 
lines of atomic N are excited by fluorescence 
and collisions with electrons and atoms in Orion. 
Because the collisional excitation of 
metastable states violates the 
assumption of downward transitions in the 
calculation of the cascade matrix, 
the population densities of the metastable and ground 
states, $n_j$ and $n_k$, 
must be solved for from equations of the type: 
\beq
n_j\sum_{i\neq j}(A_{ji}+C_{ji})=\sum_{k\neq j}n_k(A_{kj}+C_{kj}+
\beta^{\rm eff}_{kj})
\label{metast}
\eeq
where $A_{ji}=0$ if state $i$ is energetically  higher than $j$, 
and $C_{ji}$ is the collisional excitation or deexcitation rate 
from states $j$ to $i$. 
The summations comprise all the metastable and ground states, 
except state $j$. 
The system of equations~(\ref{metast}) is non--linear 
because the effective fluorescence rates depend on the 
optical depth of transitions from the ground and metastable states 
to excited states (see equations~[2] through [7] of EM).  
We found that the system~(\ref{metast}) can be solved iteratively by 
updating the optical depths across the nebula, 
and then using matrix inversion to solve for the 
populations $n_j$ at each point in the nebula in each iteration.  
The five--state inversion formulae by 
\citet{kafatos} can be easily modified to 
include the fluorescence rates in \ecua{metast}.  
The emissivity of a forbidden line is then
\beq
\epsilon_{ji}^{\rm f}=h\nu A_{ji}n_j
\label{emismetast}
\eeq

Effective recombination coefficients depend on temperature 
and more weakly on density. 
They have been tabulated with their radiative and dielectronic 
parts for spectroscopic terms of astrophysically 
important ions in the literature. 
We assume that \alfaef\ for an individual fine structure 
state of a given term is proportional to the 
relative weight of the state $(2J+1)/\sum_i(2J_i+1)$. 
This is a good approximation for direct recombinations 
from the continuum. 
Nevertheless it is worth noticing that \alfaef\ adds the 
contribution of a large number of cascading bound--bound 
transitions after the free-bound transition, 
and probabilities of fine structure transitions 
are not proportional to relative weights. 
To date only the calculation of effective recombination 
coefficients of \np\ by FSL has considered the 
fine structure splitting of quantum states. 

In the calculation of \betaef\ 
we split quantum states in their fine structure 
components either assuming 
the LS or LK 
angular momentum coupling schemes as appropriate or 
using published intermediate coupling calculations 
when available. 
The fine structure of states can be important in fluorescence 
calculations because quantum selection rules can 
selectively pump certain states. 

The calculation of \betaef\ involves 
a large number of quantum states that can be reached 
by downward transitions following an absorption from the 
ground or a metastable state. 
Metastable states and some core--excited states of the form 
\term{2s2p^{m+1}} can build up a significant 
population under low density conditions because 
of their low transition probabilities to the 
ground state. 
The population densities, $n_g$, in 
\ecua{emis} include all the fine structure states of 
the ground and metastable terms. 
Optical lines involving core--excited configurations have 
often been attributed to low temperature dielectronic 
recombination \citep{nussb81}, 
but radiative recombination or fluorescence followed by 
two--electron transitions can also produce 
excited--core configurations. 
EM found that it is important to take into account 
spin forbidden transitions and core excited configurations 
in the calculation of $\beta^{\rm eff}_{gj}$. 

\section{Atomic data}
\label{atdata}

We used the compilation in the NIST Atomic Spectra Database
\citep*{nist} to classify quantum states and store the observed
energies and probabilities for some transitions.
However we relied substantially on more recent and complete
calculations of probabilities from other authors. 
Systematic variations among different atomic databases
have small effects in the cascade matrix in most cases because 
the calculation depends on branching ratios.

\subsection{\cp}

This ion has the simplest structure. 
The ground core term produces a doublet manifold of 
excited states \term{2s^2(^2S)\ nl\ ^2L}. 
Consequently free--bound transitions tend to concentrate 
in fewer states, 
and give rise to more intense lines from states with 
higher quantum numbers. 
The excited cores \term{2s2p}, 
and \term{2p^2} produce quartet and doublet manifolds. 
We used the populations of the fine structure states of 
terms \term{2s^22p\ ^2P^o} and \term{2s2p^2\ ^4P} as the 
variables $n_g$ in \ecua{emis} for this ion. 
Some excited core configurations, 
like \term{2s2p(^3P^o)\ 3s} and \term{2s2p(^3P^o)\ 3p}, 
were also detected in \ic\ by SWBH. 

We used term--averaged transition probabilities calculated 
by \citet{nahar95} with the R--matrix approach supplemented 
with spin forbidden transitions from the NIST database. 
The latter compiles the Opacity Project calculations by 
\citet*{yan}, 
and some intermediate coupling calculations by \citet{nussb81}. 
We split transition probabilities among fine structure components 
assuming pure LS coupling \citep{allen} since no significant departures 
from that coupling have been reported \citep*{wiese}. 
We used the effective recombination coefficients calculated 
by \citet*{davey}. 

\subsection{\np}

The ground--core term of this ion 
produces singlet and triplet manifolds of 
excited states. 
The fine--structure populations of the terms 
\term{2s^22p^2\ ^3P}, \term{^1D} and \term{^1S}, 
were used as the populations $n_g$ in \ecua{emis}. 
Terms with the excited cores \term{2s2p^2\ ^4P} 
and \term{^2D} were detected in \ic\ by SWBH. 
Those cores produce singlet, triplet and quintet manifolds. 
The quintets are not reachable by transitions from the continuum 
if LS coupling selection rules prevail. 
Therefore the lines observed by SWBH between \lala5526.2 and 
5551.9, 
which are probably identified with the \nii\ multiplet 
\mult{2s2p^2(^4P)3s\ ^5P}{2s2p^2(^4P)3p\ ^5D^o}, 
may be produced by dielectronic recombination 
out of LS coupling followed by stabilizing transitions to 
the \term{3p\ ^5D^o} term. 
Recombination coefficients for those  
transitions have not been calculated yet to our knowledge. 

The atomic data to calculate this spectrum was 
described in detail by EM. 
Different calculations of transition probabilities 
---with due account of the fine structure splitting and breakdown 
of the LS coupling scheme in 4f and 5g states--- 
generally agree \citep*{lavin}. 

\begin{table*}
 \centering
  \caption{Comparison of effective recombination coefficients 
   and $A$ values for selected \nii\ lines at $10^4\K$. 
(1) Branching ratio of the line or multiplet from EV, 
(2) statistical weight of term or state, 
(3) FSL, 
(4) EV, 
(5) \citet{mar}, 
(6) \citet{shen}, 
(7) \citet{victor}.}
  \label{fangesc}
  \begin{tabular}{@{}llrrrrr@{\hspace{25pt}}rrr}
  \hline
  Transition&\multicolumn{1}{c}{$\lam$(\AA)}&\multicolumn{1}{c}{$P_{ji}$}&$g$&
\multicolumn{3}{c}{$\alfaji(10^{-15}\cccs)$}&
\multicolumn{3}{c}{$A_{ji}(10^8\persecond)$}\\ 
&&\multicolumn{1}{c}{(1)}&\multicolumn{1}{c}{(2)}&
\multicolumn{1}{r}{(3)}&\multicolumn{1}{r}{(3)}&
\multicolumn{1}{c}{(4)}&
\multicolumn{1}{c}{(5)}&\multicolumn{1}{c}{(6)}&\multicolumn{1}{c}{(7)}\\
&&&&$n_e=10^3\percubcm$&$10^4\percubcm$\\
  \hline
\mult{3p\ ^3D}{3d\ ^3F^o}&5006    &0.942$^\dagger$&21&&&238.00$^\dagger$ \\
\mult{3p\ ^3D_3}{3d\ ^3F^o_4}&5005.15&1.000&9&77.20&105.00&108.28&---&---&1.246\\
\mult{3p\ ^3D_3}{3d\ ^3F^o_3}&5025.66&0.092&7&8.69&7.92&7.78&0.118&---&0.107\\
\mult{3p\ ^3D_2}{3d\ ^3F^o_3}&5001.47&0.907&7&85.60&78.00&76.36&---&---&1.050\\
\mult{3p\ ^3D_2}{3d\ ^3F^o_2}&5016.30&0.140&5&9.60&7.43&8.42&0.186&---&0.162\\
\mult{3p\ ^3D_1}{3d\ ^3F^o_2}&5001.13&0.844&5&58.20&45.00&50.75&---&---&0.976\\
\\
\mult{3d\ ^3P^o}{4f\ D}&4433    &0.416$^\dagger$&20&&&27.20$^\dagger$ \\
\mult{3d\ ^3P^o_0}{4f\ D[3/2]_1}$^{\mbox{\S}}$&4433.48*&0.461&3&---&---&4.52&---&1.300&1.070 \\
\mult{3d\ ^3P^o_1}{4f\ D[5/2]_2}&4442.02&0.258&5&5.17&5.35&4.22&0.695&0.824&0.578 \\
\mult{3d\ ^3P^o_2}{4f\ D[5/2]_3}&4432.74*&0.830&7&18.40&19.00&18.98&---&1.790&1.926 \\
\\
\mult{3d\ ^1P^o}{4f\ D}&4690    &0.129$^\dagger$&20&&&8.46$^\dagger$ \\
\mult{3d\ ^1P^o_1}{4f\ D[3/2]_2}&4678.14*&0.317&5&---&---&5.19&---&1.030&0.717 \\
\mult{3d\ ^1P^o_1}{4f\ D[5/2]_2}&4694.64&0.481&5&6.18&6.40&7.88&0.607&0.753&1.076 \\
\\
\mult{3d\ ^1D^o}{4f\ F}&4174    &0.182$^\dagger$&28&&&18.30$^\dagger$ \\
\mult{3d\ ^1D^o_2}{4f\ F[7/2]_3}&4171.60&0.508&7&16.50&12.10&12.76&0.448&0.544&1.181\\
\mult{3d\ ^1D^o_2}{4f\ F[5/2]_3}&4176.16*&0.380&7&13.20&9.18&9.54&1.130&1.280&0.886 \\
\\
\mult{3d\ ^3D^o}{4f\ F}&4242    &0.554$^\dagger$&28&&&55.60$^\dagger$ \\
\mult{3d\ ^3D^o_1}{4f\ F[5/2]_2}&4236.93*&0.747&5&21.00&15.00&13.39&---&1.850&1.758 \\
\mult{3d\ ^3D^o_3}{4f\ F[5/2]_2}&4246.86?&0.004&5&---&---&0.07&---&0.007&0.009 \\
\mult{3d\ ^3D^o_2}{4f\ F[5/2]_3}&4241.76*&0.456&7&20.60&14.40&11.43&---&0.728&0.886 \\
\mult{3d\ ^3D^o_3}{4f\ F[7/2]_4}&4241.79*&0.889&9&36.10&26.80&28.69&---&1.960&2.090 \\
\mult{3d\ ^3D^o_2}{4f\ F[7/2]_3}&4237.05?&0.343&9&38.58&6.29&8.60&---&0.986&0.797\\
\\
\mult{3d\ ^3F^o}{4f\ F}&4089    &0.069$^\dagger$&28&&&6.92$^\dagger$ \\
\mult{3d\ ^3F^o_4}{4f\ F[7/2]_4}&4095.90&0.104&9&2.77&2.05&3.35&---&0.144&0.244 \\
\mult{3d\ ^3F^o_4}{4f\ F[7/2]_3}&4096.57&0.004&7&0.103&0.075&0.10&---&---&0.009\\
\mult{3d\ ^3F^o_3}{4f\ F[7/2]_4}&4082.27&0.007&9&6.86&5.08&0.22&0.335&0.485&0.016 \\
\mult{3d\ ^3F^o_3}{4f\ F[7/2]_3}&4082.89&0.043&7&0.41&0.30&1.08&---&0.011&0.100 \\
\mult{3d\ ^3F^o_2}{4f\ F[7/2]_3}&4073.05&0.004&7&7.90&5.81&0.096&0.499&0.631&0.0089 \\
\\
\mult{3d\ ^1F^o}{4f\ G}&4543    &0.234$^\dagger$&32&&&28.30$^\dagger$ \\
\mult{3d\ ^1F^o_3}{4f\ G[9/2]_4}&4530.41*&0.535&9&15.30&19.90&18.18&1.450&1.490&1.242\\ 
\mult{3d\ ^1F^o_3}{4f\ G[7/2]_4}&4552.52*&0.424&9&9.10&8.53&14.41&0.611&0.650&0.993 \\
\\
\mult{3d\ ^3F^o}{4f\ G}&4039    &0.765$^\dagger$&32&&&92.40$^\dagger$ \\
\mult{3d\ ^3F^o_3}{4f\ G[9/2]_4}&4026.08&0.436&9&7.15&9.34&14.81&0.672&0.794&1.013 \\
\mult{3d\ ^3F^o_4}{4f\ G[9/2]_5}&4041.31*&0.999&11&28.60&37.90&41.48&2.080&2.780&2.430 \\
\mult{3d\ ^3F^o_3}{4f\ G[7/2]_4}&4043.53*&0.540&9&18.00&16.90&18.34&1.250&1.130&1.266 \\
\mult{3d\ ^3F^o_2}{4f\ G[7/2]_3}&4035.08*&0.917&7&19.10&17.90&24.24&1.300&1.650&2.232 \\
\mult{3d\ ^3F^o_4}{4f\ G[7/2]_3}&4058.16?&0.0012&7&0.080&0.074&0.033&---&---&0.0031 \\

  \hline

\end{tabular}

\medskip

\begin{flushleft}
$^{\mbox{\S}}$ The notation in LK coupling is 4f $[K]_J$.\\
{}* Observed in \ic.\\ 
? Uncertain identification in \ic.\\
$^\dagger$ \alfaji averaged over $K$ values.\\
\end{flushleft}

\end{table*}

Most calculations of effective recombination coefficients 
give rates for terms in some coupling scheme, 
usually LS coupling. 
\citet{escvict} (hereafter EV) calculated \alfaji\ for the f states 
assuming an LK coupling scheme. 
In that scheme, the core total spin, $S_p=1/2$ is added to the 
total angular momentum $L'$ to produce an intermediate 
angular momentum $K=\vert S_p-L'\vert\ldots\vert S_p+L'\vert$ 
with values 3/2, 5/2, 7/2 and 9/2 for the f states. 
The valence electron spin is added to $K$ to give a doublet 
of total angular momentum $J=K\pm1/2$ for each LK term. 
\citet{victor} and EV give transition 
probabilities and \alfaji\ averaged over the $K$ states of 
each $m{\rm f}\ L'$ term. 
To get the transition probability to a given $K$ term, 
the $K$--averaged probability of the transition 
$n{\rm d}\ ^{1,3}L\to m{\rm f}\ L'$ must 
be multiplied by the relative weight $(2K+1)/(2S_p+1)(2L'+1)$ 
of the term. 
The probability of the inverse transition 
$m{\rm f}\ L' \to n{\rm d}\ ^{1,3}L$ is the same 
for all the $K$ states of the $m{\rm f}\ L'$ term.
We split the $3{\rm d}\ ^{1,3}L$--$4{\rm f}\ L'$ multiplets 
into fine structure components with formulae given by \citet{escgong}. 
Effective recombination coefficients of LK terms were 
multiplied by the relative weight, $(2J+1)/(2S_p+1)(2L'+1)$, 
to get the fine structure \alfaef. 
All of the above calculations to obtain the recombination and 
transition rates to individual fine structure states assume 
that the radial part of 
the dipole transition element of an LS or LK term does not change 
significantly among different $J$ states of a term. 

Recently FSL calculated effective recombination  
coefficients of \nii\ fine structure states in intermediate coupling 
and including the effects of dielectronic recombination.  
Their rates of certain lines show a hitherto unknown 
strong dependence on electron density due to the variation 
of the populations of the parent ion ground states with density. 
Table~\ref{fangesc} shows a comparison of effective recombination 
coefficients at $10^4\K$ and two electron densities for some of the 
most intense lines excited by recombination at that temperature. 

There is general agreement between the new effective recombination 
coefficients of FSL and those of EV at the electron densities of 
$n_e=10^3\percubcm$ and $10^4\percubcm$, 
but there are significant 
differences in some transitions, most notably in multiplet 
\mult{3d\ ^3F}{4f\ F[7/2]}. 
Since there is agreement between the two calculations in some 
transitions of that term, 
the differences in the other transitions with the same 
upper state appear to arise from the branching ratio of 
the transition, $P_{ji}$, or more specifically from the transition 
probability. 
For the transitions with the largest disagreements in \alfaji, 
the probabilities of \citet{victor} are much smaller than the 
experimental measurements by \citet{mar} and relativistic 
calculations by \citet*{shen}.  
Those discrepancies may point to important relativistic effects 
in some transitions that were not taken into account in the 
term--averaged values of \citet{victor} and the way in 
which we split the probabilities among the fine structure 
components. 
Unfortunately those transitions were not detected in \ic\ 
by SWBH, 
although they should be just below their detection limit 
if the larger branching ratios are correct. 

\subsection{\op}

This ion has a ground configuration \term{2s^22p^3}. 
The populations of the fine--structure states of its terms 
\term{^4S^o}, \term{ ^2D^o}\ and \term{^2P^o}\ were 
used in \ecua{emis} as the $n_g$ variables. 
Those terms are connected to quartet and doublet manifolds 
of excited states with several cores. 
SWBH possibly identified 85 transitions with the 
\term{2s^22p^2\ ^3P} ground core,  
32 transitions with the \term{2s^22p^2\ ^1D} excited core, 
12 transitions involving both cores, 
and a few lines with the \term{2s2p^3\ ^5S^o} excited core. 

We used the transition probabilities of \citet{nahar10} 
supplemented with the intermediate coupling calculations for the 
\mult{3d}{4f} array by \citet{liu95}, who found significant 
effects of LS coupling breakdown in 4f and some 3d terms. 
Terms with $l\ge3$ are better described in LK coupling 
\citep{wenaker}. Effective recombination coefficients 
were taken from \citet{liu95} for the 3d and 4f terms, 
and from \citet{storey} for other states. 
The effective recombination coefficients of \citet{nussb84} 
for terms with the \term{2s^22p^2\ ^1D} core involve only the 
dielectronic contribution. 
Transitions between states with the \term{2s^22p^2\ ^3P} 
core and states with the \term{2s^22p^2\ ^1D} core, 
some observed with non--negligible intensity in \ic, 
may add a contribution of radiative recombination 
to the terms with the \term{2s^22p^2\ ^1D} core 
at lower temperatures. 

\subsection{N and O}

For nitrogen we used the NIST compilation of transition 
probabilities mentioned at the beginning of this section. 
For oxygen we used the model potential calculations 
of \citet{escvict94} complemented with the NIST compilation 
for spin forbidden and two--electron transitions. 
Since these atoms are abundant in the neutral zone, 
the main transitions pumping the lines by fluorescence 
will be below the Lyman limit. 
Effective recombination coefficients were taken from 
\citet{escvict92} and \citet*{pequignot}. 
Effective collision strengths for excitation of 
O metastable states were taken 
from \citet{berrington} and \citet*{bell98} for electron 
collisions and \citet*{krems} and \citet*{abrahamsson} 
for collisions with H atoms. 
Effective collision strengths for excitation of N metastable 
states by electrons were taken from \citet{tayal}. 
Transition probabilities for the N and O metastable states 
were taken from \citet{charlotte}.

Atomic N has the same atomic configurations as \op\ and 
the populations of the five states of the metastable terms 
\term{^4S^o}, \term{ ^2D^o} and \term{^2P^o} were found from 
\ecua{metast}. 
Atomic O has a ground configuration \term{2s^22p^4} with 
terms \term{^3P}, \term{^1D} and \term{^1S}, 
and its excited states in the ground core configuration 
form quintet and triplet manifolds. 
There is a singlet manifold with a \term{2s^22p^3(^2D^o)} 
excited core. 
SWBH detected 10 possible lines with that core, 
but there are no calculations of effective recombination 
coefficients for them. 
Recombination from the ground states of \op\ or 
fluorescence from the \term{^3P} states can affect 
the populations of the metastable singlet states \term{^1D} 
and \term{^1S} through spin--forbidden transitions from the 
excited triplet states. 
Those transitions will be important in the cascade only 
if the stronger resonant transitions of the triplets 
are optically thick, the so called case~B \citep{escvict92}, 
and the temperature is low enough to prevent 
collisions with electrons and H atoms from 
dominating the populations of the O metastable states. 

\section{Model calculations}

The efficiency of fluorescence to enhance
dipole--allowed lines depends on the transfer
of starlight, the intensity of the diffuse
field and the ion distribution throughout the nebula.
Therefore a detailed SED and model of the nebula are required to
predict adequately the intensity of permitted lines.

\subsection{Stellar energy distribution}
\label{stellar}

\begin{table}
\caption{Ions used in the stellar model}
\label{tab_atoms}
\begin{tabular}{@{}lrr@{}}
\hline
Ions & Levels & Superlevels \\
\hline
H\thinspace{\sc i}    &  30   &   20  \\
He\thinspace{\sc i}   &  69   &   45  \\
He\thinspace{\sc ii}  &  30   &   30  \\
C\thinspace{\sc ii}   &  92   &   40  \\
C\thinspace{\sc iii}  &  243  &   99  \\
C\thinspace{\sc iv}   &  64   &   59  \\
N\thinspace{\sc ii}   &  85   &   45  \\
N\thinspace{\sc iii}  &  70   &   34  \\
N\thinspace{\sc iv}   &  76   &   44  \\
N\thinspace{\sc v}    &  67   &   45  \\
O\thinspace{\sc ii}   &  123  &   54  \\
O\thinspace{\sc iii}  &  170  &   88  \\
O\thinspace{\sc iv}   &  78   &   38  \\
O\thinspace{\sc v}    &  152  &   75  \\
O\thinspace{\sc vi}   &  13   &   13  \\
Si\thinspace{\sc iii} &  34   &   20  \\
Si\thinspace{\sc iv}  &  33   &   22  \\
P\thinspace{\sc iv}   &  178  &   36  \\
P\thinspace{\sc v}    &  62   &   16  \\
S\thinspace{\sc iii}  &  28   &   13  \\
S\thinspace{\sc iv}   &  142  &   51  \\
S\thinspace{\sc v}    &  98   &   31  \\
S\thinspace{\sc vi}   &  58   &   28  \\
Ne\thinspace{\sc ii}  &  48   &   14  \\
Ne\thinspace{\sc iii} &  71   &   23  \\
Ne\thinspace{\sc iv}  &  52   &   17  \\
Ar\thinspace{\sc iii} &  36   &   10  \\
Ar\thinspace{\sc iv}  &  105  &   31  \\
Ar\thinspace{\sc v}   &  99   &   38  \\
Fe\thinspace{\sc iii} &  1433 &   104 \\
Fe\thinspace{\sc iv}  &  1000 &   100 \\
Fe\thinspace{\sc v}   &  1000 &   139 \\
Fe\thinspace{\sc vi}  &  433  &   44  \\
Fe\thinspace{\sc vii} &  153  &   29  \\
\hline
\end{tabular}
\end{table}

\begin{figure*}
 \includegraphics[width=18.cm]{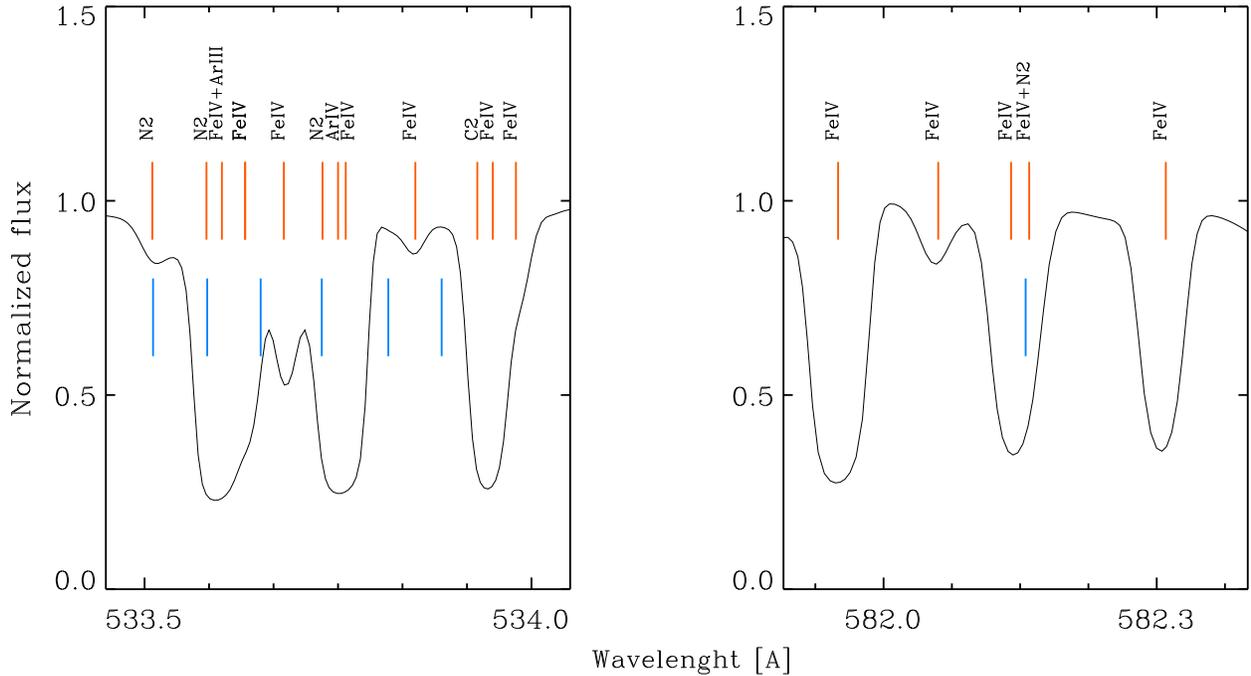}
 \caption{Continuum--normalized SED with identifications of 
  stellar lines. 
  Unlabelled lower vertical lines (blue) show the wavelengths 
  of the transitions in \nii\ multiplet   
  \mult{2p^2\ ^3P}{3d\ ^3D^o}\lala533.51--533.88, 
  and \nii\ line \mult{2p^2\ ^1D_2}{3d\ ^1D^o_2}\lam582.156 
  in the rest frame of the photosphere.} 
 \label{sed}
\end{figure*}

To evaluate the amount of emitted energy better 
and increase the resolution of the SED near the wavelengths 
of the most important pumping transitions, 
we calculated a new model of the central star atmosphere using the 
\cmf\ code (Hillier and Miller, 1998) with a more complex 
atomic structure than the one used in MG as shown in 
table~\ref{tab_atoms}. 
The increased number of lines in the far UV 
region of the spectrum changed the ionization balance of the nebula.
Following the same procedure used by MG, 
we adjusted the stellar atmosphere model,
increasing its temperature to 39000\K\ and keeping the other 
parameters unchanged. 
The new self--consistent stellar and nebular models 
reproduce the intensities of the emission lines 
common to both models with a quality factor $\kappa(O)$ 
(see definition in equation~(6) of MG) of the same order of 
magnitude as the one obtained with the previous SED. 
The new high--dispersion ($R\sim100 000$) SED 
was corrected for a stellar rotation velocity of 
30\kms\ as derived in MG.

The Doppler shift of the resonant transition frequencies 
due to internal motions of the nebular gas 
must be taken into consideration with a SED of that resolution 
as discussed in section~\ref{velocity}. 
We adopted a velocity field $V\propto R^4$ with a 
maximum velocity $V_m=40\kms$ at the outer edge of the nebula. 
MG adopted a similar velocity law to reproduce the profiles 
of the \hb, [\nii] and [\oiii] lines (see appendix~B of MG). 
The effects of other velocity fields and of turbulence on 
the calculated line intensities are discussed in section~\ref{velocity}. 

We did not find a systematic decrease of fluorescence 
excitation of the CNO lines in the SWBH survey 
with a higher--dispersion SED as reported for the 
Balmer lines by \citet{vale}. 
The probable reason is the fact that 
excited states in multielectron species have several 
resonant transitions distributed over a range of wavelengths.  
The fluorescence excitation to those states  
can increase or decrease with a 
high--dispersion SED depending on the coincidence of those  
resonant transitions with the emission peaks or absorption 
troughs of the stellar wind profile in the rest frame of 
the absorbing ion or atom. 
The resolution of the adopted SED is larger than the 
width of the pumping lines in the gas. 
Therefore we do not expect any remaining bias in the 
predicted intensities with the adopted SED. 

Fluorescence excitation depends on a large number of 
resonant transitions, 
but there are a few transitions especially important 
in the populations of certain states. 
Fig.~\ref{sed} shows two sections of the SED 
with examples of important transitions in the excitation 
of \nii\ lines by fluorescence. 
The low abundance of \np\ in the stellar atmosphere 
produces little depression of the continuum at the 
\nii\ pumping transitions in the gas, 
but other species do produce important absorption 
features in the SED as shown in Fig.~\ref{sed}

\subsection{Photoionization model}

The photoionization model calculated by MG took into 
account the small ellipticity of \ic\ by using the 
pseudo--3D code \clotd\ \citep{chris06}, 
which allows the calculation of nebular models with 
complex geometries. 
The near spherical symmetry of \ic\ and the 
small angle of the aperture used by SBW 
allowed us to approximate the nebula with a 1D spherical  
model to speed up the calculation of thousands of 
transitions in this work. 
We extracted continuum opacities, ground and metastable 
state populations, ion and electron densities, and 
electron temperatures from version c08.00 of \clo 
\citep{cloudy} with the SED described above and a 
density distribution similar to the one used by MG. 
Emissivities calculated with equations~(\ref{emis}) 
and~(\ref{emismetast}) are then 
integrated along a beam across the nebula to simulate 
the size and position of the aperture used by SBW. 
Images of the slit position on the nebula are given in MG.

The stellar light is attenuated by geometrical dilution and 
continuum opacity as given by \clo. 
The self--absorption within the resonant transition is 
taken into account with the `pumping probability' defined by 
\citet{ferland}. 
The calculation of \betaef\ uses the escape probability formalism 
of \citet{humstorey} to approximate the propagation of photons in 
the resonant transitions with an overlapping continuum in the nebula. 
This in effect takes into account the variation of optical depth 
in a medium partially thick in the resonant transitions,  
a situation named `case D' by \citet{vale}. 
Additional details can be found in EM. 

Although MG used some nebular lines to adjust the stellar 
atmosphere and nebular model parameters, 
most of their calculated and all of the calculated 
line intensities in the present work are {\it ab initio} predictions 
of the SED and nebular model. 
The abundances adjusted by MG with this procedure and used in this 
work are given in table~\ref{abunds}. 
The density profile in this work is 
\begin{eqnarray}
n(R<R_{\rm in})&=&0 \nonumber\\
n(R_3>R>R_{\rm in})&=&n_0+n_1\exp\big[(R-R_1)^2/\sigma_1^2\big]+ \\
&&n_2\exp\big[(R-R_2)^2/\sigma_2^2\big]\nonumber
\label{perfil}
\end{eqnarray}
where $R$ is the distance to the central star in cm. 
Table~\ref{params} gives the adopted values for 
parameters $n_{0\ldots2}$, $R_{\rm in}$, $R_1$, $R_2$,  
$\sigma_1$ and $\sigma_2$. 
Parameters $R_1$, $R_2$ have intermediate 
values between the long and short axes in the ellipsoidal 
model of MG.
This density profile was chosen to match the HST 
images of the intense nebular lines as discussed in detail 
by MG (see their figure~5). 
With few exceptions, the predicted line intensities 
in our 1D model vary by less than 30 per cent when 
$n_{0\ldots2}$ are varied simultaneously by 20 per cent or 
when $R_1$ and $R_2$ are varied simultaneously by 20 per cent. 

\begin{table}
\caption{Element abundances in the nebular model}
\label{abunds}
\begin{tabular}{@{}lr@{}}
\hline
Element &$\log {\rm X/H}$\\
\hline
He &$-0.92$\\
C &$-3.10$\\
N &$-4.00$\\
O &$-3.40$\\
Ne &$-4.00$\\
Mg &$-4.95$\\
Si &$-4.90$\\
S &$-5.35$\\ 
Cl &$-6.90$\\
Ar &$-5.80$\\
Fe &$-7.40$\\
\hline

\end{tabular}

\end{table}

\begin{table}
\caption{Nebular density profile parameters}
\label{params}
\begin{tabular}{@{}@{$}l@{$}lr@{}}
\hline
\hbox{Parameter }&Unit&Value \\
\hline
n_0 &\percubcm& 2850 \\
n_1 &\percubcm& 5260 \\
n_2 &\percubcm& 9500 \\
n_3 &\percubcm& 20500 \\
\log R_{\rm in} &cm& 16.06 \\
\log R_1  &cm& 16.28 \\
\log R_2  &cm& 17.02 \\
\log R_3  &cm& 17.10 \\   
\log \sigma_1  &cm& 15.96 \\
\log \sigma_2  &cm& 16.46 \\
\hline

\end{tabular}

\end{table}

\subsection{The photodissociation region}
It has long been known that \ic\ has a 
PhotoDissociation Region (PDR) 
extending beyond the optical nebula \citep{cohen,taylor87} 
with low ionization and atomic components 
\citep*{monk,taylor89,meixner}. 
\citet{taylor89} measured an H mass of $0.35\pm0.05\,$M\sun\  
over a radius $\ga90\,$arcsec from the center of the nebula. 
Thus far there have been no detections of molecular emission in 
\ic\ \citep{huggins,dayal,hora}. 
\citet{liu01a} obtained an average density for the PDR of 
$10^{5.5}\percubcm$ from observations of the [\cii] 158--$\,\mu$m 
and [\oi] 63--$\,\mu$m and 146--$\,\mu$m lines.  
That density is much higher than the one needed to 
explain the radio observations and suggests the 
existence of a narrow dense shell just outside the 
ionization front or clumps in the PDR. 

The \clo\ nebular model can be readily extended to the PDR 
by adding an outer neutral atomic envelope to the ionized 
region starting at a distance of the central star $R_3$. 
We chose the following density profile for the PDR: 
\beq
n_{\rm PDR}(R > R_3)=n_3 (T_e/10^4\K)^{-1.2} 
\label{pdr}
\eeq
where $T_e$ is the electron temperature in kelvins,  
and $n_3$ and $R_3$ are given in table~\ref{params}. 
This density profile is constrained by the observed fluxes 
of the FIR [\cii] and [\oi] lines. 
The temperature dependence in Eq.~(\ref{pdr}) allows for 
a smooth transition between the ionized and neutral material 
by increasing the density as the temperature drops. 
The extent of the neutral region needed to 
match the observed intensities of the FIR \oi\ and \cii\ 
lines is just a few percent of the radius of the ionized region. 
A more realistic model of the ionized and molecular 
interface needs a hydrodynamic model, which is 
beyond the scope of this paper.

\section{Results}

\subsection{Selection of observations}
\label{observs}

Proper line identification of weak spectral lines 
is a fundamental problem in the analysis 
of deep nebular spectroscopy. 
SWBH presented the line identification code \emi\ to 
automate and standarise an otherwise tedious labour of 
identifying several hundreds of spectral lines.  
SBW selected a subset of emission line intensities 
from the survey of SWBH based on the likelihood of a  
correct, unique identification as given by \emi\ 
and the availability of published effective 
recombination coefficients to determine abundances. 
Some lines in SWBH were not taken into account by SBW because 
those lines were probably blended with other emission lines or 
\emi\ gave more than one plausible identification for them. 

We have reexamined the SWBH line list and added 
many measured lines of \cii, \ni, \nii, \oi\ and \oii\ 
to the selection of SBW in order to compare them with 
our calculations either because we could estimate 
their effective recombination coefficients reasonably 
well or because they were the most probable contributor 
to a blend. 
Some cases for individual species are discussed in the 
following sections. 

Published calculations often tabulate the line effective 
recombination coefficient in \ecua{alfaef} 
for the most intense lines only. 
Coefficients for weaker lines with the same upper state $j$ 
can be readily estimated by dividing by the branching ratio 
$P_{ji}$ of the tabulated transition and multiplying 
by the corresponding $P_{ji}$ of the desired line as in 
\ecua{emis}. 
In order to do this consistently, it is necessary that 
authors publish the transition probability data or at least the 
branching ratios used in their calculations 
\citep{nussb84, escvict, pequignot, escvict92, liu95}. 

The issue of multiple identifications is more complicated.  
A line with multiple identifications may be either a real blend 
or some of the proposed identifications may not be real. 
Our calculations show that in many cases a transition 
in a possible blend contributes insignificantly to the 
measured intensity. 
When a transition in a blend of two transitions of 
the same species contributes more than 10 per cent 
of the total intensity, 
we have added the calculated intensities of the transitions 
to compare them with the measurements by SWBH. 
Blends of emissions of the same species are thus compared with our 
computed spectra without the need to assume theoretical 
intensities to `deblend' the observed lines. 
Blends of emissions of different species are compared 
directly with the predicted intensities.

\subsection{Comparison with observed intensities}

\begin{figure*}
 \includegraphics[width=18.0cm,bb=18 288 592 574,clip]{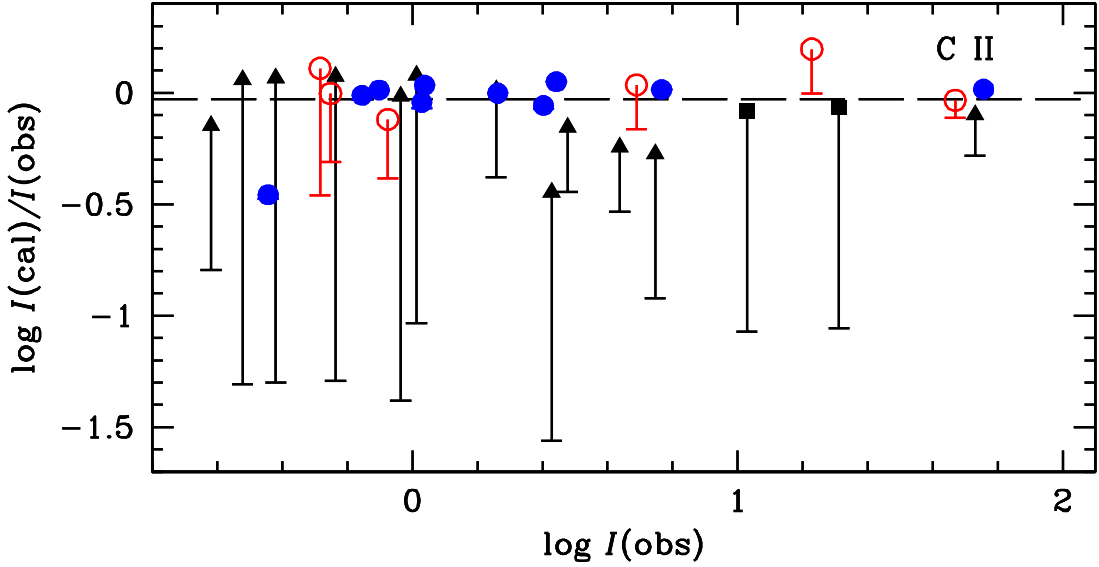}
 \caption{Comparison of \cii\ intensities observed by SBW 
  with predicted intensities produced by recombination and fluorescence 
  of lines from s states (black squares), 
  p states (black triangles), 
  d states (open red circles), 
  and f and g states (filled blue circles).  
  The dash at the bottom of the vertical lines marks the intensity produced 
  by recombination excitation only. 
  The broken horizontal line is the average value of
  $I_{\rm cal}/I_{\rm obs}$.
  Note that the observed intensities 
  are normalized to $I(\hb)=10^4$.} 
 \label{ciicompar}
\end{figure*}

\begin{figure*}
 \includegraphics[width=18.0cm,bb=18 288 592 574,clip]{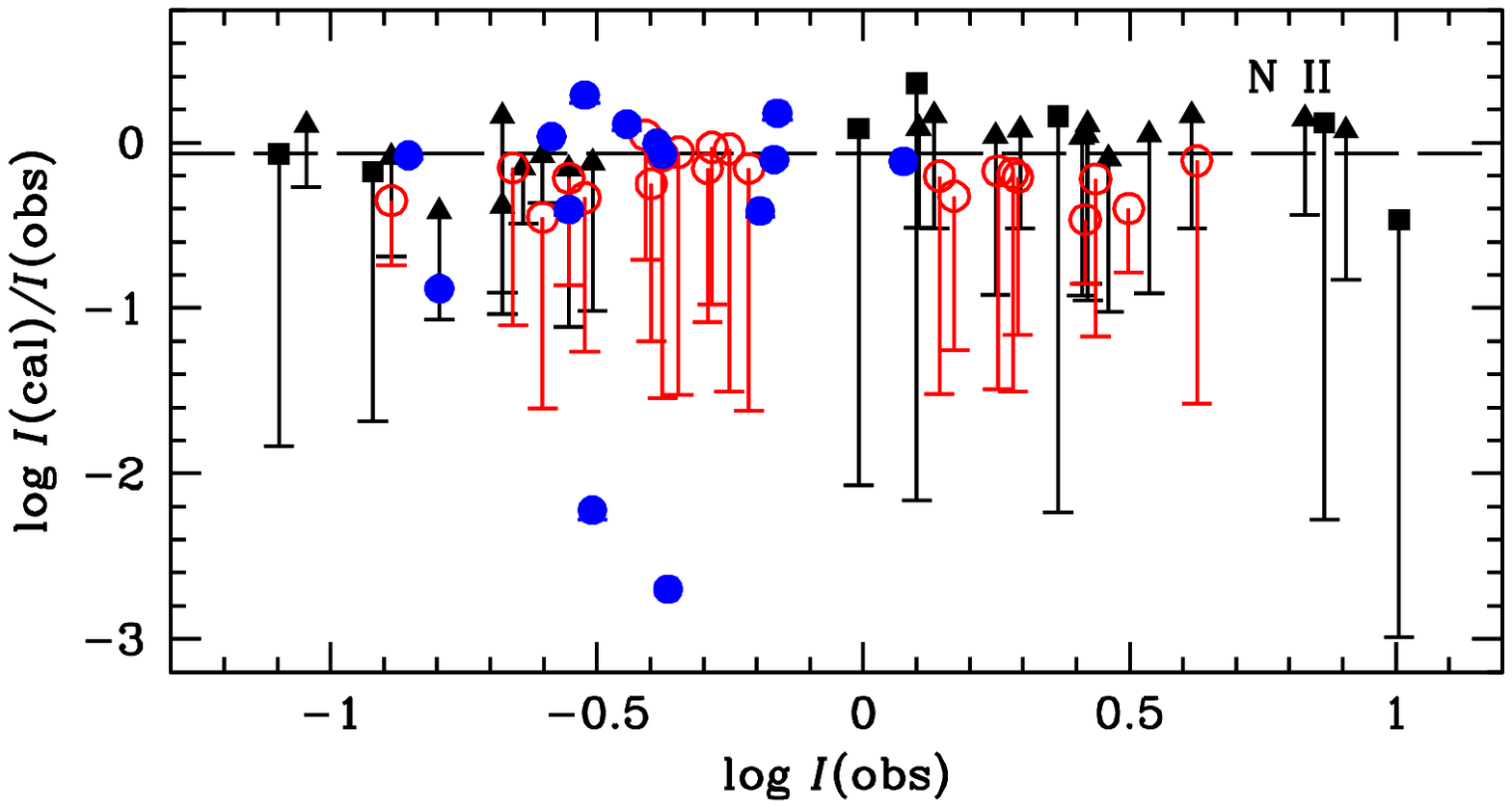}
 \caption{Same as in Fig.~\ref{ciicompar} for \nii\ lines.} 
 \label{niicompar}
\end{figure*}

\begin{figure*}
 \includegraphics[width=18.0cm,bb=18 288 592 574,clip]{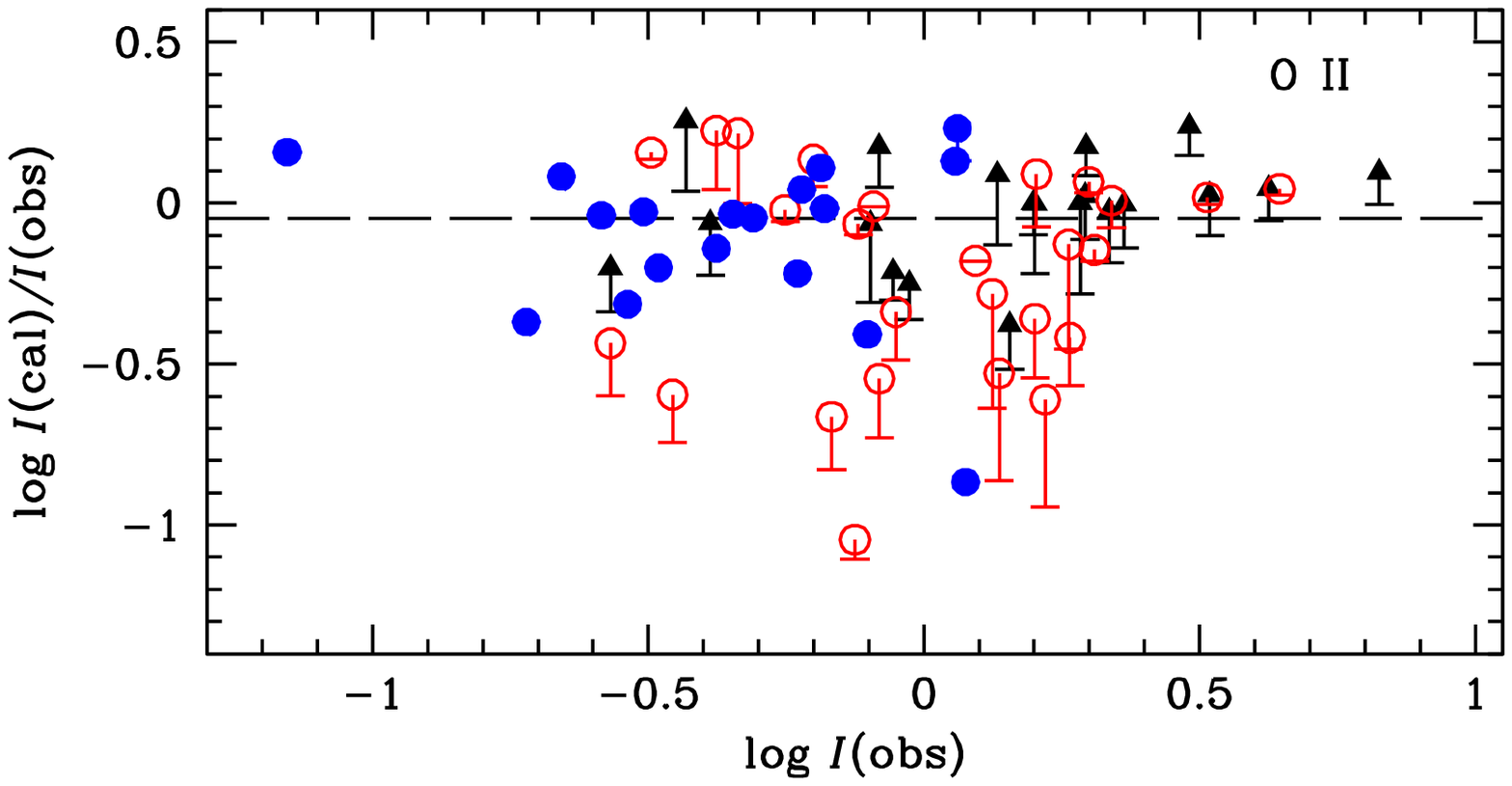}
 \caption{Same as in Fig.~\ref{ciicompar} for \oii\ lines.} 
 \label{oiicompar}
\end{figure*}

\begin{figure*}
 \includegraphics[width=18.0cm,bb=18 288 592 574,clip]{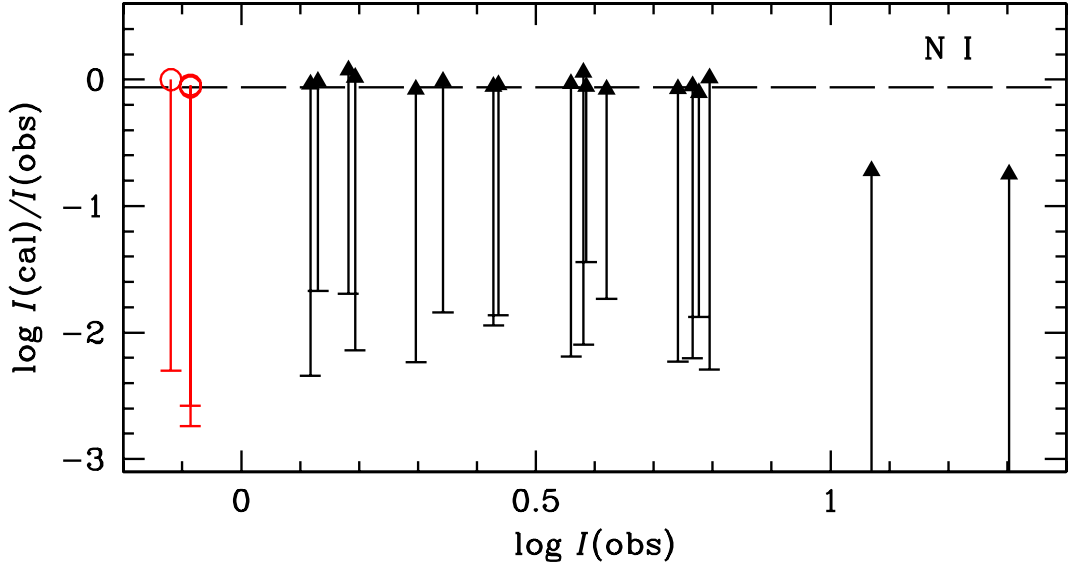}
 \caption{Same as in Fig.~\ref{ciicompar} for \ni\ lines. 
  Collisionally excited forbidden lines are also included.} 
 \label{nicompar}
\end{figure*}

\begin{figure*}
 \includegraphics[width=18.0cm,bb=18 288 592 574,clip]{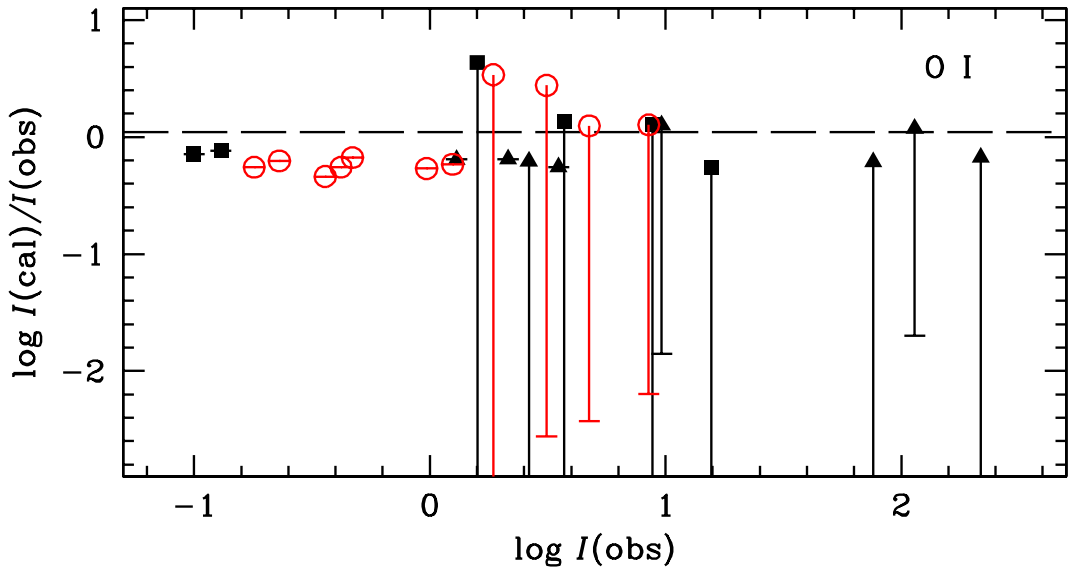}
 \caption{Same as in Fig.~\ref{nicompar} for \oi\ lines.} 
 \label{oicompar}
\end{figure*}

In Fig.~\ref{ciicompar} to~\ref{oicompar} we compare 
predicted intensities with observed intensities in a 
scale of $I(\hb)=10^4$. 
Tables~\ref{tabcii} to~\ref{taboi} give the 
predicted intensities of dipole--allowed 
lines with intensities higher than 
$8\times10^{-6}$ of the \hb\ strength, 
which is close to the lower limit of detection in SWBH. 
Tables~\ref{tabni} and~\ref{taboi} also include predicted 
intensities of the optical forbidden lines of \ni\ and \oi. 
Some of the wavelengths given in the second column of 
the table are calculated from energy levels and have 
limited accuracy. 
The fraction of calculated intensity due to recombination is 
given in the third column of the tables. 
The calculated and observed intensities in the 
fourth and fifth columns are normalized to 
$I(\hb)=10^4$. 
The bottom line in each table gives the average of 
the ratio of the intensity due to recombination to 
the total calculated intensity and the average of 
the ratio of the calculated intensity to the observed 
intensity for each species. 
We have found some undetected lines 
with calculated intensities above the 
detection limit in SWBH. 
Some of them are blended with intense nebular lines, 
and a few were measured, but apparently not identified by \emi.
The observed intensity of lines not selected by SBW 
are marked with `?' in the sixth column of the tables. 

The uncertainty of each line intensity depends on several 
factors like the signal--to--noise ratio ($\rm S/N$), its 
relative width with respect to the instrument resolution 
and whether the line may be blended with other lines. 
In order to make meaningful comparisons with observations, 
MG estimated a variable uncertainty of the 
line fluxes in SBW between 10 per cent for the brighter lines to 
30 per cent for the weaker lines. 
SBW suggest an uncertainty of 20 per cent for lines with 
a $\rm S/N>20$. 
Most of the lines analysed in this work 
are weaker than 0.01 times the intensity of \hb, 
and have a $\rm S/N>20$. 
Therefore we will adopt a minimum general uncertainty of 
20 per cent for all dipole--allowed lines in the SWBH survey. 

Our calculations tend to underestimate 
most of the lines with dubious identification 
according to \emi. 
\citet{rola} and \citet*{wesson} have shown that measured intensities 
of lines with $\rm S/N<6$ are strongly biased upwardly. 
We only have two lines with an underestimated predicted intensity 
and a measured $\rm S/N<7$, which corresponds to a true 
$\rm S/N$ between 6 and 8 at the 1--$\sigma$ confidence level 
according to the calculations of \citet{rola}. 
In most of our cases it is more likely that the underestimated 
predictions indicate incorrect identifications or 
blends with other lines rather than uncertainties in 
the measured values. 
We have not included in the tables and figures 
features with a dubious identification and a 
predicted intensity less than 0.1 of the measured value. 

The excitation mechanisms of a spectrum can be 
diagnosed by contrasting the intensities of lines 
yielded by different atomic configurations 
as done by \citet{grandi75,grandi76}. 
Lines from s, p and most d states with the 
same spin multiplicity as that of the ground state are 
enhanced in varying degrees by fluorescence relative 
to recombination lines 
if the recombined ion concentration is sufficiently high. 
Transitions from d states that are not connected to the ground 
or a metastable state by a resonant transition and transitions 
from f and g states are excited mostly or totally by recombination. 

SBW found that some dipole--allowed lines of a species 
have profiles similar to those of the forbidden lines of 
the next higher ionization stage of the species, 
(e.g.~some \oi\ line profiles are similar to [\oii] profiles), 
indicating that they are produced in the same part of the 
nebula by recombination of the more ionized stage.  
Likewise they found that other dipole--allowed lines of the 
same species have profiles similar to those of the 
forbidden lines of the same ionization stage, indicating 
that they are produced by fluorescence excitation of that species. 

SBW also found a correlation between the line width of the 
species producing the line and the ionization potential of 
the species, 
which indicates a relationship between the line width and 
the ionization stratification in the nebula. 
Recombination lines of species with lower ionization 
potential generally have broader lines and their widths are similar 
to those of the forbidden lines of the next higher ionization stage, 
indicating again a common place of origin within the nebula. 

SBW proposed that line profile and line width similarities 
can be used to discriminate the excitation mechanism of the lines. 
The contributions of fluorescence and recombination, however, 
are comparable in many lines and produce many exceptions 
to the discriminating criteria proposed by SBW as discussed for 
particular cases below. 
Therefore a more quantitative analysis is needed to determine 
the excitation mechanism of a line. 

\subsection{\cii}

The agreement of predicted recombination intensities within 
20 per cent of observed values in the SBW selection 
of 9 lines or blends of lines from f and g states supports  
the accuracy of the recombination rates for this ion. 
That remarkable consistency can be explained by three 
factors concerning the observed high angular 
momentum states: 
those states are less dependent on atomic calculation assumptions 
because of their reduced non--hydrogenic effects,  
their populations are free from optical depth effects, 
and their fine--structure components are often blended 
in spectroscopic terms. 
In those blends, the effective transition rate of the 
term is a weighted average of several fine structure 
transition probabilities so that the total intensity of 
the blend is the sum of intensities of those components. 
Quantum calculations with different assumptions tend to 
agree on the transition probabilities for the more 
intense components of a term. 
The average tends to cancel out differences in the 
calculated $A$--values of the fine structure components 
while the summation gives more weight to the components 
with the larger intensities and lower observational errors. 
Measurement errors are also expected to decrease 
with the higher S/N of a blend of several lines belonging 
to the same term. 

Recombinations produce more than half the 
intensity of most \cii\ lines, 
but we find an important fluorescence contribution to 
all the observed lines from s, p and d states.  
SBW did not find differences in the line profiles 
of those lines with profiles of recombination lines as should be 
indicated by their proposed criteria based on 
line profile differentiation. 
The fluorescence contribution is particularly important in 
multiplets \mult{3p\ ^2P^o}{4s\ ^2S_{1/2}}\lala3918.97, 3920.68, 
\mult{4p\ ^2P^o}{6d\ ^2D}\lala4637.63--4639.07, 
\mult{3d\ ^2D}{4p\ ^2P^o}\lala5889.28--5891.60, 
\mult{4p\ ^2P^o}{5d\ ^2D}\lala 6257.18--6259.80, 
and \mult{3s\ ^2S_{1/2}}{3p\ ^2P^o}\lala6578.05, 6582.88, 
and it is mostly due to absorptions from the ground state 
in the doublet \mult{2p\ ^2P^o}{4s\ ^2S}\lala636.994, 636.251, 
and the triplets \mult{2p\ ^2P^o}{5d\ ^2D}\lala560.239--560.439, 
\mult{2p\ ^2P^o}{6d\ ^2D}\lala543.258--543.445, 
followed by decays to term \term{4p\ ^2P^o}. 
This fluorescence excitation explains the enhanced intensities 
of those multiplets relative to the recombination lines, 
and the seemingly high \cpp\ abundances derived from their 
intensities by SBW using recombination rates alone. 

We added 13 lines from the SWBH survey with excited--core 
upper terms \term{2s2p(^3P^o)3p\ ^2P} and \term{2s2p(^3P^o)3p\ ^2D} 
and two lines from the \term{9f\ ^2F^o} term 
using additional effective recombination coefficients of 
\citet{davey} published on line at the CDS site. 
Term \term{2s2p(^3P^o)3p\ ^2P} is mostly excited by fluorescence 
through absorptions in multiplet 
\mult{2p\ ^2P^o}{2s2p(^3P^o)3p\ ^2P}\lala549.320--549.570. 
The calculated intensities for decays from that term agree with  
the observations within our adopted uncertainty intervals except 
for the line \mult{2p^3\ ^2P^o_{1/2}}{2s2p(^3P^o)3p\ ^2P_{3/2}}\lam7508.89 
apparently undetected by SWBH, and the line 
\mult{4p\ ^2P_{3/2}}{2s2p(^3P^o)3p\ ^2P_{3/2}}\lam5121.83 
underestimated by a factor of 2.8. 
Lines from term \term{2s2p(^3P^o)3p\ ^2D} on the other hand 
are underestimated by our calculations by 30 to 50 per cent.
Lines from the state \term{2s2p(^3P^o)3p\ ^2D_{3/2}} have a 
70 per cent excitation by fluorescence while lines 
from the other state \term{2s2p(^3P^o)3p\ ^2D_{5/2}} have 
a 40 per cent excitation by fluorescence mainly in absorptions in 
multiplet \mult{2p\ ^2P^o}{2s2p(^3P^o)3p\ ^2D}\lala530.275--530.474. 
The relatively intense line 
\mult{4p\ ^2P^o_{1/2}}{2s2p(^3P^o)3p\ ^2D_{3/2}}\lam3835.72 is 
probably blended with the Balmer H9 line. 
The two blended lines 
\mult{5d\ ^2D_{5/2}}{9f\ ^2F^o_{5/2}} and 
\mult{5d\ ^2D_{5/2}}{9f\ ^2F^o_{7/2}} at \lam7860.50 are  
the only f multiplet with a strong discrepancy with observations  
with an observed intensity 2.8 times larger than the predicted 
intensity. 
The discrepancy may be due to uncertainties in the measurement 
because the observed S/N of 28.7 in SWBH is lower than the 
other \cii\ blends of f and g states and the intensity of the other 
blend from the \term{9f\ ^2F} term is within the uncertainty interval 
of the observed value. 

\begin{table}
\caption{Predicted and observed line intensities of \cii. 
$I_{\rm rec}/I_{\rm calc}$ is the fraction of the calculated 
intensity due to recombination, 
$I_{\rm calc}$ is the predicted intensity, $I_{\rm obs}$ is the 
measured intensity with a 20 per cent minimum uncertainty. The ? 
marks measured intensities taken from SWBH, but not selected by SBW.}
\label{tabcii}
\begin{tabular}{llclll}
\hline
Lower--Upper&\lam(\AA)&$I_{\rm rec}/I_{\rm calc}$&$I_{\rm calc}$&$I_{\rm obs}$\\
\hline
\mult{4p\ ^2P^o_{3/2}}{3p'\ ^2D_{5/2}}$\dagger$& 3831.7&0.512& 2.10& 3.00 ? \\ 
\mult{4p\ ^2P^o_{1/2}}{3p'\ ^2D_{3/2}}& 3835.7&0.226& 2.51&--- \\ 
\mult{4p\ ^2P^o_{3/2}}{3p'\ ^2D_{3/2}}& 3836.7&0.226& 0.50&--- \\ 
\mult{3p\ ^2P^o_{1/2}}{4s\ ^2S_{1/2}}& 3919.0&0.102& 8.87&10.68   \\ 
\mult{3p\ ^2P^o_{3/2}}{4s\ ^2S_{1/2}}& 3920.7&0.102&17.70&20.52   \\ 
\mult{3d\ ^2D_{5/2}}{4f\ ^2F^o_{7/2}}& 4267.3&0.989&59.20&57.12   \\ 
\mult{4f\ ^2F^o}{10g\ ^2G} \S& 4292.2&1.000& 0.81& 0.79   \\ 
\mult{4d\ ^2D}{9f\ ^2F^o}& 4329.7&0.954& 0.68& 0.70   \\ 
\mult{4f\ ^2F^o}{9g\ ^2G}& 4491.1&1.000& 1.18& 1.09   \\ 
\mult{4d\ ^2D}{8f\ ^2F^o}& 4618.8&0.942& 0.97& 1.07   \\ 
\mult{4p\ ^2P^o_{1/2}}{6d\ ^2D_{3/2}}& 4637.6&0.492& 0.56& 0.56   \\ 
\mult{4p\ ^2P^o_{3/2}}{6d\ ^2D_{5/2}}& 4638.9&0.807& 0.65&--- \\ 
\mult{4p\ ^2P^o_{3/2}}{6d\ ^2D_{3/2}}& 4639.1&0.492& 0.11&--- \\ 
\mult{2p^2\ ^2P_{3/2}}{3p\ ^2P^o_{3/2}}& 4744.8&0.654& 0.09&--- \\ 
\mult{4f\ ^2F^o}{8g\ ^2G}& 4801.9&1.000& 1.82& 1.83   \\ 
\mult{2p^3\ ^2P^o_{3/2}}{3p'\ ^2D_{5/2}}& 5032.1&0.512& 2.48& 4.34 ? \\ 
\mult{2p^3\ ^2P^o_{1/2}}{3p'\ ^2D_{3/2}}& 5035.9&0.226& 2.97& 5.58 ? \\ 
\mult{2p^3\ ^2P^o_{3/2}}{3p'\ ^2D_{3/2}}& 5040.7&0.226& 0.59&--- \\ 
\mult{4p\ ^2P^o_{1/2}}{3p'\ ^2P_{3/2}}& 5120.1&0.077& 0.19&--- \\ 
\mult{4p\ ^2P^o_{3/2}}{3p'\ ^2P_{3/2}}& 5121.8&0.077& 0.96& 2.68 ? \\ 
\mult{4p\ ^2P^o_{1/2}}{3p'\ ^2P_{1/2}}& 5125.2&0.043& 0.69& 0.58 ? \\ 
\mult{4p\ ^2P^o_{3/2}}{3p'\ ^2P_{1/2}}& 5127.0&0.043& 0.34& 0.30 ? \\ 
\mult{4f\ ^2F^o}{7g\ ^2G}& 5341.5&1.000& 3.11& 2.77   \\ 
\mult{3d\ ^2D_{3/2}}{4p\ ^2P^o_{3/2}}& 5889.3&0.506& 0.30&--- \\ 
\mult{3d\ ^2D_{5/2}}{4p\ ^2P^o_{3/2}}& 5889.8&0.506& 2.69&--- \\ 
\mult{3d\ ^2D_{3/2}}{4p\ ^2P^o_{1/2}}& 5891.6&0.406& 1.86& 1.81   \\ 
\mult{4d\ ^2D}{6f\ ^2F^o}& 6145.9&0.969& 2.22& 2.53   \\ 
\mult{4p\ ^2P^o_{1/2}}{5d\ ^2D_{3/2}}& 6257.2&0.270& 0.67& 0.52   \\ 
\mult{4p\ ^2P^o_{3/2}}{5d\ ^2D_{5/2}}& 6259.6&0.544& 0.64& 0.84   \\ 
\mult{4p\ ^2P^o_{3/2}}{5d\ ^2D_{3/2}}& 6259.8&0.270& 0.13&--- \\ 
\mult{4f\ ^2F^o}{6g\ ^2G}& 6460.1&1.000& 6.04& 5.84   \\ 
\mult{3s\ ^2S_{1/2}}{3p\ ^2P^o_{3/2}}& 6578.1&0.654&42.90&53.74   \\ 
\mult{3s\ ^2S_{1/2}}{3p\ ^2P^o_{1/2}}& 6582.9&0.527&26.60&--- \\ 
\mult{3p\ ^2P^o_{1/2}}{3d\ ^2D_{3/2}}& 7231.3&0.632&26.60&16.92   \\ 
\mult{3p\ ^2P^o_{3/2}}{3d\ ^2D_{5/2}}& 7236.4&0.836&43.30&46.73   \\ 
\mult{3p\ ^2P^o_{3/2}}{3d\ ^2D_{3/2}}& 7237.2&0.632& 5.30& 4.89   \\ 
\mult{5f\ ^2F^o_{5/2}}{10g\ ^2G_{7/2}}& 7508.9&1.000& 0.17&--- \\ 
\mult{5f\ ^2F^o_{7/2}}{10g\ ^2G_{9/2}}& 7508.9&1.000& 0.23&--- \\ 
\mult{2p^3\ ^2P^o_{1/2}}{3p'\ ^2P_{3/2}}& 7508.9&0.077& 0.25&--- \\ 
\mult{2p^3\ ^2P^o_{3/2}}{3p'\ ^2P_{3/2}}& 7519.5&0.077& 1.23& 1.03 ? \\ 
\mult{2p^3\ ^2P^o_{1/2}}{3p'\ ^2P_{1/2}}& 7519.9&0.043& 0.89& 0.92 ? \\ 
\mult{2p^3\ ^2P^o_{3/2}}{3p'\ ^2P_{1/2}}& 7530.6&0.043& 0.44& 0.38 ? \\ 
\mult{5d\ ^2D_{3/2}}{9f\ ^2F^o_{5/2}}& 7860.2&0.943& 0.09&--- \\ 
\mult{5d\ ^2D_{5/2}}{9f\ ^2F^o_{5/2}}& 7860.5*&0.962& 0.13& 0.36 ? \\ 
\mult{5f\ ^2F^o_{5/2}}{9g\ ^2G_{7/2}}& 8139.6&1.000& 0.24&--- \\ 
\mult{5f\ ^2F^o_{7/2}}{9g\ ^2G_{9/2}}& 8139.6&1.000& 0.31&--- \\ 
\mult{5d\ ^2D_{3/2}}{8f\ ^2F^o_{5/2}}& 8867.6&0.923& 0.11&--- \\ 
\mult{5d\ ^2D_{5/2}}{8f\ ^2F^o_{7/2}}& 8868.1&0.956& 0.16&--- \\ 
\hline
\end{tabular}

\S\ Terms with no subscript are blended transitions of several states 
of angular momentum $J$.\\
$\dagger$ 3p' is core excited configuration \term{2s2p(^1D^o)3p}.\\
{}*\lam7860.5 includes blend with \mult{5d\ ^2D_{5/2}}{9f\ ^2F^o_{7/2}}.
\end{table}
\begin{table}
\contcaption{}
\begin{tabular}{llclll}
\hline
Lower--Upper&\lam(\AA)&$I_{\rm rec}/I_{\rm calc}$&$I_{\rm calc}$&$I_{\rm obs}$\\
\hline
\mult{5f\ ^2F^o_{7/2}}{8g\ ^2G_{9/2}}& 9221.1&1.000& 0.45&--- \\ 
\mult{5f\ ^2F^o_{5/2}}{8g\ ^2G_{7/2}}& 9221.1&1.000& 0.35&--- \\ 
\mult{3s'\ ^2P^o_{3/2}}{3p'\ ^2D_{5/2}}& 9238.3&0.512& 0.14&--- \\ 
\mult{3s'\ ^2P^o_{1/2}}{3p'\ ^2D_{3/2}}& 9251.0&0.226& 0.17& 0.24 ? \\ 
\hline
\multicolumn{2}{r}{$\langle I_{\rm rec}/I_{\rm calc}\rangle=0.581$}
&\multicolumn{3}{r}{$\langle I_{\rm calc}/I_{\rm obs}\rangle=0.938$}\\
\hline
\end{tabular}

\end{table}

\subsection{\nii}
\label{sectnii}

Fluorescence is the main excitation mechanism 
for most \nii\ lines in \ic, 
but to a lesser extent than in the Orion nebula. 
The main pumping mechanisms in this ion were discussed 
in detail by \citet{orion}. 
The \term{^3P} ground core offers several pumping 
channels to the \term{3d\ ^3P^o} and \term{3d\ ^3D^o} 
terms through multiplets 
\mult{2p^2\ ^3P}{3d\ ^3P^o}\lala529.36--529.87, 
\mult{2p^2\ ^3P}{3d\ ^3D^o}\lala533.51--533.88 and 
\mult{2p^2\ ^3P}{4s\ ^3P^o}\lala508.48--509.01. 
The only 3d term with significant recombination excitation 
is \term{3d\ ^3F^o}, 
but its brightest components at \lala5001.13, 5001.13 and 
5005.15 are blended with the [\oiii]\lam5006.84 
nebular line. 
The singlet lines \mult{3p\ ^1P_1}{3d\ ^1D^o_2}\lam4447.03 and 
\mult{3p\ ^1D_2}{3d\ ^1F^o_3}\lam6610.56 listed by SBW are 
also excited mostly by fluorescence due to absorptions in the 
transitions 
\mult{2p\ ^3P_1}{3d\ ^1D^o_2}\lam534.657 and
\mult{2p\ ^1D_2}{3d\ ^1F^o_3}\lam574.650 respectively. 
The first one is a spin--forbidden transition with the 
ground triplet term while the second demonstrates that  
a metastable state can also produce fluorescence excitation. 

The disagreement of observed intensities with predicted 
intensities in lines excited by fluorescence is 
greater in this ion. 
There is a systematic underestimation of 
the calculated values of lines from 3d states of 30 to 70 per cent. 
The rest of the calculated intensities selected by SBW 
are within 40 per cent of the observed values. 

The 4f states are populated by recombination. 
Direct recombinations and transitions from the 4f states 
produce between 30 and 60 per cent of the intensity 
of the lines from the \term{3d\ ^3D^o} and \term{3d\ ^3F^o} 
terms. 
EM showed that fluorescence cannot significantly populate 
f and higher angular momentum states in \np\ 
because quantum probabilities strongly favour 
downward transitions with decreasing angular momentum. 
That is a well known trend of transition probabilities 
inferred from atomic data compilations 
and theoretical considerations \citep{rusos}. 
Consequently d states have a greater probability of decaying to 
p states than to f states. 
Should fluorescence enhance an f state population 
by decays from a higher d state, lines from lower p states,  
like \mult{3s\ ^3P^o_2}{3p\ ^3P_2}\lam4630.54, 
would show at least a 70 per cent enhancement in their 
predicted intensity (EM). 
In our present calculations that line is already 
overpredicted by 20 per cent above its observed value without 
any plausible fluorescence contribution to f states. 
The enhanced intensity of the  
\mult{3s\ ^3P^o_2}{3p\ ^3D_3}\lam5679.56 line by 
fluorescence in \ic\ puts its ratio with the recombination line 
\mult{3d\ ^3F^o_4}{4f\ G[9/2]_5}\lam4041.31 out of 
the $T_e$ and $n_e$ ranges in fig.~8 of FSL, 
and highlights the importance of determining the 
atomic processes that excite a line before it is 
used as a diagnostic tool. 

We have added 15 lines from the SWBH survey that were 
omitted in the analysis of SBW.  
Some like \mult{3d\ ^3D^o_2}{4f\ F[5/2]_3}\lam4241.76,  
and \mult{3d\ ^3D^o_3}{4f\ F[7/2]_4}\lam4241.78 are 
blended, 
and \mult{3d\ ^3D^o_2}{4f\ F[7/2]_3}\lam4237.05 is probably 
blended with the transition 
\mult{3d\ ^3D^o_1}{4f\ F[5/2]_2}\lam4236.91 listed in the 
selection of SBW. 
There are four lines standing out from f states with predicted 
intensities significantly below the observed values: 
\mult{3d\ ^3F^o_4}{4f\ G[7/2]_3}\lam4058.16, 
\mult{3d\ ^1D^o_2}{4f\ F[5/2]_3}\lam4176.16, 
\mult{3d\ ^3D^o_3}{4f\ F[5/2]_2}\lam4246.71, 
\mult{3d\ ^3P^o_0}{4f\ D[3/2]_1}\lam4433.48,  
and \mult{4d\ ^1F_3}{5f\ G[9/2]_4}\lam9794.05, 
which are discussed below. 

Because there is only one more line from a 4f~D term 
with a secure identification 
(\mult{3d\ ^1P^o_1}{4f\ D[3/2]_2}\lam4678.14) 
and an intensity 15 per cent above the predicted value,  
we cannot assert whether 
the effective recombination coefficient of the \lam4433.48 line 
is underestimated. 
The other line produced by a 4f~D term listed in SWBH, 
\mult{3d\ ^3P^o_2}{4f\ D[5/2]_3}\lam4432.74, 
may be blended with a [Fe~II] line, 
but our calculations predict an intensity 30 per cent above 
the measured value. 
The effective recombination coefficients of FSL 
and EV at $10^4\K$ agree for the lines of multiplet 
\mult{3d\ ^3P^o}{4f\ D[5/2]}\lala4432.74--4442.02.  

Our calculations predict intensities 
0.002 times and 0.006 times the observed intensities of 
the lines \lam4058.16 and \lam4246.71 respectively. 
The low S/N of 10.3 and 9.2 
and the displacement between the tabulated and observed 
wavelength of $10.3\kms$ and $9.2\kms$ respectively reported by SWBH 
for those lines suggest that they may be misidentifications. 
Other lines of the term \term{4f\ F[5/2]}, 
like \mult{3d\ ^1D^o_2}{4f\ F[5/2]_3}\lam4176.16, with 
more secure identifications also show important differences with 
the calculated values. 

The dispersion of calculated intensities above and below the observed 
values of transitions from 4f~D and 4f~F terms 
may imply uncertainties in the recombination coefficients or 
uncertainties in the measurements due to the low signal to noise ratio.
There are no significant differences between the 
effective recombination coefficients of FSL and EV 
for those lines as shown in table~\ref{fangesc}. 

For the lines from 4f~G terms the agreement between predicted and 
observed intensities is within 20 per cent for nearly all lines. 
The only exception is the line \mult{3d\ ^3F^o_4}{4f\ G[7/2]_3}
\lam4058.16 mentioned above 
with an effective recombination coefficient by FSL 
more than twice the value of EV. 
Even so, the value by FSL is not large enough to explain 
its intensity if it is indeed \nii\ emission. 

Finally the line \mult{4d\ ^1F_3}{5f\ G[9/2]_4}\lam9794.05 
is the only line from a \term{5f\ G} term above the detection 
limit, 
and is strongly underestimated by a factor of 7.6 in intensity,  
with a S/N of 9.0 and no alternative identifications. 

The line at \lam4674.903 was measured but not identified by SWBH. 
Our calculations predict the \nii\ spin--forbidden transition 
\mult{3s\ ^1P^o_1}{3p\ ^3P_0}\lam4674.91 to be a 
most likely identification. 
The transition probability given by~\citet{charlotte} 
of $8.54\times10^6\persecond$ overestimates 
the observed value by 40 per cent. 
The other two members of this \nii\ multiplet,  
\mult{3s\ ^1P^o_1}{3p\ ^3P_2}\lam4654.53 and 
\mult{3s\ ^1P^o_1}{3p\ ^3P_1}\lam4667.21, were 
identified by SWBH as the \oi\  
multiplet \mult{3p\ ^5P}{8d\ ^5D^o}\lam4654.56 and 
the line [\mbox{Fe\,{\sc iii}}]
\mult{3d^6\ ^5D_3}{3d^6\ ^3F2_2}\lam4667.010 respectively. 
Our calculations predict an intensity 0.09 times weaker than the 
observed value for the \oi\ multiplet (see section~\ref{sectoi}) while 
the agreement with predicted intensities for both lines is much 
better if our proposed identifications as \nii\ lines are correct. 

\begin{table}
\caption{Predicted and observed line intensities of \nii.}
\label{tabnii}
\begin{tabular}{llclll}
\hline
Lower--Upper&\lam(\AA)&$I_{\rm rec}/I_{\rm calc}$&$I_{\rm calc}$&$I_{\rm obs}$\\
\hline
\mult{3p\ ^3S_1}{4s\ ^3P^o_2}& 3593.6&0.004& 1.79&--- \\ 
\mult{3p\ ^3S_1}{4s\ ^3P^o_1}& 3609.1&0.003& 1.41&--- \\ 
\mult{3p\ ^3S_1}{4s\ ^3P^o_0}& 3615.9&0.007& 0.22&--- \\ 
\mult{3p\ ^3P_1}{4s\ ^3P^o_2}& 3829.8&0.004& 3.36& 2.32   \\ 
\mult{3p\ ^3P_2}{4s\ ^3P^o_2}& 3838.4&0.004& 9.68& 7.34 ? \\ 
\mult{3p\ ^3P_0}{4s\ ^3P^o_1}& 3842.2&0.003& 2.87& 1.26   \\ 
\mult{3p\ ^3P_1}{4s\ ^3P^o_1}& 3847.4&0.003& 2.08&--- \\ 
\mult{3p\ ^3P_1}{4s\ ^3P^o_0}& 3855.1&0.007& 1.19& 0.98   \\ 
\mult{3p\ ^3P_2}{4s\ ^3P^o_1}& 3856.1&0.003& 3.47&10.13 ? \\ 
\mult{3d\ ^3F^o_3}{4f\ G[9/2]_4}& 4026.0&0.999& 0.33&--- \\ 
\mult{3d\ ^3F^o_2}{4f\ G[7/2]_3}& 4035.0&1.000& 0.54& 0.68   \\ 
\mult{3d\ ^3F^o_4}{4f\ G[9/2]_5}& 4041.3&1.000& 0.92& 1.19   \\ 
\mult{3d\ ^3F^o_3}{4f\ G[7/2]_4}& 4043.5&0.999& 0.41& 0.41   \\ 
\mult{3d\ ^3F^o_4}{4f\ G[7/2]_3}& 4058.1&1.000& 0.00& 0.43 ? \\ 
\mult{3d\ ^3F^o_4}{4f\ F[7/2]_4}& 4095.9&0.910& 0.09&--- \\ 
\mult{3d\ ^1D^o_2}{4f\ F[7/2]_3}& 4171.5&0.944& 0.33&--- \\ 
\mult{3d\ ^1D^o_2}{4f\ F[5/2]_3}& 4176.1&0.927& 0.25& 0.64   \\ 
\mult{3d\ ^3D^o_3}{4f\ D[5/2]_3}& 4179.6&0.923& 0.09&--- \\ 
\mult{3d\ ^3D^o_1}{4f\ F[5/2]_2}& 4236.9*&0.900& 0.58& 0.30 ? \\ 
\mult{3d\ ^3D^o_2}{4f\ F[5/2]_3}& 4241.7*&0.915& 1.04& 0.69 ? \\ 
\mult{3d\ ^3D^o_3}{4f\ F[5/2]_2}& 4246.7&0.874& 0.00& 0.31 ? \\ 
\mult{3d\ ^3P^o_1}{4f\ D[3/2]_2}& 4427.2&0.952& 0.15&--- \\ 
\mult{3d\ ^3P^o_1}{4f\ D[3/2]_1}& 4427.9&0.930& 0.08&--- \\ 
\mult{3d\ ^3P^o_2}{4f\ D[5/2]_3}& 4432.7&0.923& 0.47& 0.36 ? \\ 
\mult{3d\ ^3P^o_0}{4f\ D[3/2]_1}& 4433.4&0.930& 0.11& 0.28   \\ 
\mult{3d\ ^3P^o_1}{4f\ D[5/2]_2}& 4442.0&0.967& 0.10&--- \\ 
\mult{3p\ ^1P_1}{3d\ ^1D^o_2}& 4447.0&0.177& 0.43& 0.39   \\ 
\mult{3p\ ^3D_1}{3d\ ^3P^o_0}& 4459.9&0.070& 0.09& 0.25   \\ 
\mult{3p\ ^3D_1}{3d\ ^3P^o_1}& 4465.5&0.034& 0.11&--- \\ 
\mult{3p\ ^3D_2}{3d\ ^3P^o_1}& 4477.7&0.034& 0.43& 0.61   \\ 
\mult{3p\ ^3D_3}{3d\ ^3P^o_2}& 4507.6&0.117& 0.36& 0.51   \\ 
\mult{3d\ ^1F^o_3}{4f\ G[9/2]_4}& 4530.3&0.999& 0.36& 0.42   \\ 
\mult{3d\ ^1F^o_3}{4f\ G[7/2]_4}& 4552.5&0.999& 0.28& 0.26   \\ 
\mult{3s\ ^3P^o_1}{3p\ ^3P_2}& 4601.5&0.126& 2.92& 2.63   \\ 
\mult{3s\ ^3P^o_0}{3p\ ^3P_1}& 4607.2&0.110& 2.79& 2.57 ? \\ 
\mult{3s\ ^3P^o_1}{3p\ ^3P_1}& 4613.9&0.110& 1.93& 1.77   \\ 
\mult{3s\ ^3P^o_1}{3p\ ^3P_0}& 4621.4&0.086& 3.42& 2.64   \\ 
\mult{3s\ ^3P^o_2}{3p\ ^3P_2}& 4630.5&0.126& 9.54& 8.05   \\ 
\mult{3s\ ^3P^o_2}{3p\ ^3P_1}& 4643.1&0.110& 3.83& 3.44   \\ 
\mult{3s\ ^1P^o_1}{3p\ ^3P_2}& 4654.5&0.126& 0.24& 0.31 ? \\ 
\mult{3s\ ^1P^o_1}{3p\ ^3P_1}& 4667.2&0.110& 0.20& 0.28 ? \\ 
\mult{3s\ ^1P^o_1}{3p\ ^3P_0}& 4674.9&0.086& 0.30& 0.21 ? \\ 
\mult{3d\ ^1P^o_1}{4f\ D[3/2]_2}& 4678.1&0.952& 0.12& 0.14   \\ 
\mult{3d\ ^1P^o_1}{4f\ D[5/2]_2}& 4694.6&0.967& 0.17&--- \\ 
\mult{3p\ ^3D_1}{3d\ ^3D^o_2}& 4774.2&0.112& 0.16& 0.22   \\ 
\mult{3p\ ^3D_1}{3d\ ^3D^o_1}& 4779.7&0.048& 1.21& 1.79   \\ 
\mult{3p\ ^3D_2}{3d\ ^3D^o_3}& 4781.2&0.409& 0.06& 0.13   \\ 
\mult{3p\ ^3D_2}{3d\ ^3D^o_2}& 4788.1&0.112& 1.20& 1.95   \\ 
\mult{3p\ ^3D_2}{3d\ ^3D^o_1}& 4793.6&0.048& 0.37&--- \\ 
\mult{3p\ ^3D_3}{3d\ ^3D^o_3}& 4803.3&0.409& 0.89& 2.61   \\ 
\mult{3p\ ^3D_3}{3d\ ^3D^o_2}& 4810.3&0.112& 0.23& 0.40 ? \\ 
\mult{3p\ ^3S_1}{3d\ ^3P^o_0}& 4987.4&0.070& 0.53&--- \\ 
\mult{3p\ ^3S_1}{3d\ ^3P^o_1}& 4994.4&0.034& 3.29& 4.23   \\ 
\mult{3p\ ^3D_1}{3d\ ^3F^o_2}& 5001.1&0.668& 1.19&--- \\ 
\mult{3p\ ^3D_2}{3d\ ^3F^o_3}& 5001.5&0.493& 2.75&--- \\ 
\mult{3s\ ^3P^o_0}{3p\ ^3S_1}& 5002.7&0.119& 0.58&--- \\ 
\mult{3p\ ^3D_3}{3d\ ^3F^o_4}& 5005.2&0.929& 2.26&--- \\ 
\mult{3p\ ^3S_1}{3d\ ^3P^o_2}& 5007.3&0.117& 2.53&--- \\ 
\mult{3s\ ^3P^o_1}{3p\ ^3S_1}& 5010.6&0.119& 1.49&--- \\ 
\mult{3p\ ^3D_2}{3d\ ^3F^o_2}& 5016.4&0.668& 0.20&--- \\ 
\mult{3p\ ^3D_3}{3d\ ^3F^o_3}& 5025.7&0.493& 0.28&--- \\ 
\hline
\end{tabular}

{}*\lam4236.9 includes blend with \mult{3d\ ^3D^o_5}{4f\ F[7/2]_7}\lam4237.047\\
\lam4241.7 includes blend with \mult{3d\ ^3D^o_7}{4f\ F[7/2]_9}\lam4241.786
\end{table}
\begin{table}
\contcaption{}
\begin{tabular}{llclll}
\hline
Lower--Upper&\lam(\AA)&$I_{\rm rec}/I_{\rm calc}$&$I_{\rm calc}$&$I_{\rm obs}$\\
\hline
\mult{3s\ ^3P^o_2}{3p\ ^3S_1}& 5045.1&0.119& 2.31& 2.89   \\ 
\mult{3s\ ^1P^o_1}{3p\ ^3S_1}& 5073.6&0.119& 0.17&--- \\ 
\mult{3p\ ^3P_0}{3d\ ^3P^o_1}& 5452.1&0.034& 0.35& 0.42   \\ 
\mult{3p\ ^3P_1}{3d\ ^3P^o_0}& 5454.2&0.070& 0.22&--- \\ 
\mult{3p\ ^3P_1}{3d\ ^3P^o_1}& 5462.6&0.034& 0.40& 0.45   \\ 
\mult{3p\ ^3P_1}{3d\ ^3P^o_2}& 5478.1&0.117& 0.14& 0.30   \\ 
\mult{3p\ ^3P_2}{3d\ ^3P^o_1}& 5480.1&0.034& 0.51& 0.56   \\ 
\mult{3p\ ^3P_2}{3d\ ^3P^o_2}& 5495.7&0.117& 0.70& 1.48   \\ 
\mult{3s\ ^3P^o_1}{3p\ ^3D_2}& 5666.6&0.208& 6.01& 4.14   \\ 
\mult{3s\ ^3P^o_0}{3p\ ^3D_1}& 5676.0&0.252& 2.37& 1.97   \\ 
\mult{3s\ ^3P^o_2}{3p\ ^3D_3}& 5679.6&0.263& 9.40& 6.74   \\ 
\mult{3s\ ^3P^o_1}{3p\ ^3D_1}& 5686.2&0.252& 1.55& 1.27   \\ 
\mult{3s\ ^3P^o_2}{3p\ ^3D_2}& 5710.8&0.208& 1.98& 1.36   \\ 
\mult{3s\ ^3P^o_2}{3p\ ^3D_1}& 5730.7&0.252& 0.11& 0.13   \\ 
\mult{3s\ ^1P^o_1}{3p\ ^3D_2}& 5747.3&0.208& 0.54&--- \\ 
\mult{3s\ ^1P^o_1}{3p\ ^3D_1}& 5767.5&0.252& 0.19&--- \\ 
\mult{3p\ ^3P_0}{3d\ ^3D^o_1}& 5927.8&0.048& 1.24& 1.91   \\ 
\mult{3p\ ^3P_1}{3d\ ^3D^o_2}& 5931.8&0.112& 1.65& 2.73   \\ 
\mult{3p\ ^3P_1}{3d\ ^3D^o_1}& 5940.2&0.048& 0.87& 1.39   \\ 
\mult{3p\ ^3P_2}{3d\ ^3D^o_3}& 5941.7&0.409& 1.26& 3.15   \\ 
\mult{3p\ ^3P_2}{3d\ ^3D^o_2}& 5952.4&0.112& 0.49& 0.52   \\ 
\mult{3d\ ^3F^o_4}{4p\ ^3D_3}& 6167.8&0.519& 0.21& 0.25   \\ 
\mult{3d\ ^3F^o_2}{4p\ ^3D_1}& 6170.2&0.423& 0.11& 0.09   \\ 
\mult{3d\ ^3F^o_3}{4p\ ^3D_2}& 6173.3&0.453& 0.16& 0.23   \\ 
\mult{3s\ ^3P^o_1}{3p\ ^1P_1}& 6379.6&0.351& 0.08&--- \\ 
\mult{3s\ ^1P^o_1}{3p\ ^1P_1}& 6482.0&0.351& 0.39&--- \\ 
\mult{3p\ ^1D_2}{3d\ ^1F^o_3}& 6610.6&0.226& 0.17& 0.28   \\ 
\mult{3d\ ^3P^o_2}{4p\ ^3S_1}& 6810.0&0.223& 0.09& 0.21   \\ 
\mult{3d\ ^3P^o_1}{4p\ ^3S_1}& 6834.1&0.223& 0.06& 0.16   \\ 
\mult{4p\ ^3D_3}{5s\ ^3P^o_2}& 8676.1&0.031& 0.08& 0.12   \\ 
\mult{4p\ ^3D_2}{5s\ ^3P^o_1}& 8699.0&0.017& 0.07& 0.08 ? \\ 
\mult{4d\ ^1F^o_3}{5f\ G[9/2]_4}& 9794.2&0.999& 0.02& 0.16 ? \\ 
\hline
\multicolumn{2}{r}{$\langle I_{\rm rec}/I_{\rm calc}\rangle=0.348$}
&\multicolumn{3}{r}{$\langle I_{\rm calc}/I_{\rm obs}\rangle=0.863$}\\
\hline

\end{tabular}
\end{table}

\subsection{\oii}

The lines in this ion are produced mostly by recombination, 
which strongly favours the quartets because of its higher 
multiplicity.  
Fluorescence is limited to excitations 
from the \term{2p^3\ ^4S^o_{3/2}} state, 
and thus contributes 
less than half the intensity of lines from the 
\term{3p\ ^4P^o} term through absorptions in the 
\mult{2p^3\ ^4S^o}{3d\ ^4P}\lala429.918--430.176 multiplet 
with some contribution from absorptions in the 
\mult{2p^3\ ^4S^o}{3d\ ^4D}\lala429.650--429.716 multiplet, 
followed by decays to the 3p terms. 
Most doublets in the ground core configuration 
are being excited by recombination with less than 
20 per cent contribution of fluorescence from the \term{2p^3\ ^2D^o} 
metastable term. 
The exception are the states of the \term{3d\ ^2F} term. 
Lines \mult{3p\ ^2D^o_{3/2}}{3d\ ^2F_{5/2}}\lam4699.22 and 
\mult{3p\ ^2D^o_{5/2}}{3d\ ^2F_{7/2}}\lam4705.35 have a 
50 per cent contribution of fluorescence from the metastable 
term \term{2p^3\ ^2D^o}. 
The spin--forbidden transition 
\mult{2p^3\ ^4S^o_{3/2}}{3d\ ^2F_{5/2}}\lam429.560  
contributes 10 per cent to the fluorescence excitation of 
the \lam4699.22 line. 
This line may be blended with the core excited transition 
\mult{2p^2(^1D)3p\ ^2D^o_{5/2}}{2p^2(^1D)3d\ ^2F_{7/2}}\lam4699.011 
for which we lack a published recombination coefficient, 
and probably contributes to the intensity of the line, 
which appears somewhat underestimated in our calculations. 

The dielectronic recombination coefficients from \citet{nussb84} 
give a good fit to the observed intensities of lines of 
configurations with the excited core \term{2p^2(^1D)}. 
We estimated the coefficient of some lines not listed 
by \citet{nussb84} by scaling the coefficients of other 
lines by the corresponding branching ratios as explained 
in section~\ref{observs}. 
Nevertheless many other \oii\ lines with an excited core listed 
in SWBH lack effective recombination coefficients and were 
left out of our calculation. 

We added eight lines identified by SWBH to the selection in SBW. 
Some may be blended with transitions from other ions, but 
\oii\  probably contributes most of the intensity of the line. 
Line \mult{3p\ ^4P^o_{1/2}}{3d\ ^4D_{3/2}}\lam4097.22 
is probably blended with line 
\mult{3d\ ^4F_{7/2}}{4f\ G[4]_{9/2}}\lam4097.26 and we have 
added their calculated intensities in table~\ref{taboii}. 

The largest discrepancy in our calculations occurs in the 
line \mult{3p\ ^4P^o_{3/2}}{3d\ ^4D_{1/2}}\lam4110.79 
with a calculated intensity a factor of 11 below 
the observed value. 
Lines with upper terms \term{3d\ ^4P} and \term{3d\ ^4D} 
also tend to be underestimated in our calculations, 
and the reason appears to be uncertainties in the 
transition probabilities. 
The $A$--values for those terms calculated by 
\citet{nahar10} are significantly lower 
than those quoted in the NIST database from 
\citet{veres} and \citet{bell}. 
The NIST values give a much better 
--albeit insufficient-- match to observations 
with predicted intensities 0.09 to 0.4 times the 
measured value, far below the assumed 
observational uncertainty. 
Some comparisons with the observed and calculated 
line intensities with different $A$--values 
are given in table~\ref{avalues}. 

\begin{table}
\caption{Predicted and observed line intensities of \oii.}
\label{taboii}
\begin{tabular}{llclll}
\hline
Lower--Upper&\lam(\AA)&$I_{\rm rec}/I_{\rm calc}$&$I_{\rm calc}$&$I_{\rm obs}$\\
\hline
\mult{3p\ ^4D^o_{1/2}}{3d\ ^4D_{3/2}}& 3842.8&0.605& 0.15&--- \\ 
\mult{3p\ ^4D^o_{3/2}}{3d\ ^4D_{5/2}}& 3850.7&0.686& 0.16&--- \\ 
\mult{3p\ ^4D^o_{3/2}}{3d\ ^4D_{3/2}}& 3851.1&0.605& 0.30&--- \\ 
\mult{3p\ ^4D^o_{5/2}}{3d\ ^4D_{7/2}}& 3863.6&0.825& 0.10&--- \\ 
\mult{3p\ ^4D^o_{5/2}}{3d\ ^4D_{5/2}}& 3864.5&0.686& 0.54&--- \\ 
\mult{3p\ ^4D^o_{5/2}}{3d\ ^4D_{3/2}}& 3864.8&0.605& 0.20&--- \\ 
\mult{3p\ ^4D^o_{7/2}}{3d\ ^4D_{7/2}}& 3882.3&0.825& 0.86& 0.63   \\ 
\mult{3p\ ^4D^o_{7/2}}{3d\ ^4D_{5/2}}& 3883.1&0.686& 0.10& 0.27   \\ 
\mult{3p\ ^4D^o_{5/2}}{3d\ ^4P_{5/2}}& 3907.6&0.711& 0.09& 0.35   \\ 
\mult{3d\ ^4F_{7/2}}{4f\ F[3]^o_{7/2}}& 4048.2&1.000& 0.10&--- \\ 
\mult{3d\ ^4F_{9/2}}{4f\ F[4]^o_{9/2}}& 4062.8&1.000& 0.19&--- \\ 
\mult{3p\ ^4D^o_{1/2}}{3d\ ^4F_{3/2}}& 4069.6&0.921& 1.46& 2.04   \\ 
\mult{3p\ ^4D^o_{3/2}}{3d\ ^4F_{5/2}}& 4069.9&0.924& 2.32& 1.99   \\ 
\mult{3d\ ^4F_{7/2}}{4f\ G[5]^o_{9/2}}& 4071.2&1.000& 0.26&--- \\ 
\mult{3p\ ^4D^o_{5/2}}{3d\ ^4F_{7/2}}& 4072.3&0.952& 3.41& 3.27   \\ 
\mult{3p\ ^4D^o_{7/2}}{3d\ ^4F_{9/2}}& 4075.9&0.956& 4.90& 4.42   \\ 
\mult{3p\ ^4D^o_{3/2}}{3d\ ^4F_{3/2}}& 4078.8&0.921& 0.53& 0.56   \\ 
\mult{3d\ ^4F_{5/2}}{4f\ G[4]^o_{7/2}}& 4083.8&1.000& 0.44& 0.49 ? \\ 
\mult{3p\ ^4D^o_{5/2}}{3d\ ^4F_{5/2}}& 4085.2&0.924& 0.65& 0.76   \\ 
\mult{3d\ ^4F_{3/2}}{4f\ G[3]^o_{5/2}}& 4087.1&1.000& 0.42& 0.45   \\ 
\mult{3d\ ^4F_{9/2}}{4f\ G[5]^o_{11/2}}& 4089.2&1.000& 1.54& 1.14   \\ 
\mult{3p\ ^4D^o_{7/2}}{3d\ ^4F_{7/2}}& 4093.0&0.952& 0.46& 0.32   \\ 
\mult{3d\ ^4F_{5/2}}{4f\ G[3]^o_{7/2}}& 4095.6&1.000& 0.30& 0.42   \\ 
\mult{3d\ ^4F_{7/2}}{4f\ G[4]^o_{9/2}}& 4097.2*&0.795& 1.96& 1.15 ? \\ 
\mult{3d\ ^4F_{3/2}}{4f\ D[3]^o_{5/2}}& 4098.2&1.000& 0.25&--- \\ 
\mult{3p\ ^4P^o_{3/2}}{3d\ ^4D_{5/2}}& 4104.8&0.686& 1.97& 1.60 ? \\ 
\mult{3p\ ^4P^o_{3/2}}{3d\ ^4D_{3/2}}& 4105.1&0.605& 0.76& 0.46   \\ 
\mult{3d\ ^4F_{5/2}}{4f\ D[3]^o_{7/2}}& 4107.0&1.000& 0.18&--- \\ 
\mult{3p\ ^4P^o_{3/2}}{3d\ ^4D_{1/2}}& 4110.9&0.868& 0.07& 0.75   \\ 
\mult{3p\ ^4P^o_{5/2}}{3d\ ^4D_{7/2}}& 4119.2&0.825& 2.23& 2.19   \\ 
\mult{3p\ ^4P^o_{5/2}}{3d\ ^4D_{5/2}}& 4120.2&0.686& 0.15& 0.68   \\ 
\mult{3p\ ^4P^o_{1/2}}{3d\ ^4P_{1/2}}& 4121.3&0.465& 0.41& 1.66   \\ 
\mult{3p\ ^4P^o_{3/2}}{3d\ ^4P_{1/2}}& 4129.3&0.465& 0.64&--- \\ 
\mult{3p\ ^4P^o_{1/2}}{3d\ ^4P_{3/2}}& 4132.7&0.655& 0.24& 0.83   \\ 
\mult{3p\ ^4P^o_{3/2}}{3d\ ^4P_{3/2}}& 4140.8&0.655& 0.42&--- \\ 
\mult{3p\ ^4P^o_{3/2}}{3d\ ^4P_{5/2}}& 4153.4&0.711& 0.70& 1.84   \\ 
\mult{3p\ ^4P^o_{5/2}}{3d\ ^4P_{3/2}}& 4156.4&0.655& 0.70& 1.59 ? \\ 
\mult{3p\ ^4P^o_{5/2}}{3d\ ^4P_{5/2}}& 4169.2&0.711& 0.96&--- \\ 
\mult{3p'\ ^2F^o_{5/2}}{3d'\ ^2G_{7/2}}$\dagger$& 4185.5&0.997& 0.79& 0.81   \\ 
\mult{3p'\ ^2F^o_{7/2}}{3d'\ ^2G_{9/2}}& 4189.9&0.997& 0.82& 1.24   \\ 
\mult{3d\ ^4D_{7/2}}{4f\ F[4]^o_{9/2}}& 4275.5&1.000& 0.84& 0.65   \\ 
\mult{3d\ ^4D_{3/2}}{4f\ F[3]^o_{5/2}}& 4275.9&1.000& 0.16&--- \\ 
\mult{3d\ ^4D_{5/2}}{4f\ F[3]^o_{5/2}}& 4276.3&1.000& 0.12&--- \\ 
\mult{3d\ ^4D_{5/2}}{4f\ F[3]^o_{7/2}}& 4276.8&1.000& 0.31& 0.79 ? \\ 
\mult{3d\ ^4D_{1/2}}{4f\ F[2]^o_{3/2}}& 4277.4&1.000& 0.21& 0.33   \\ 
\mult{3d\ ^4D_{7/2}}{4f\ F[3]^o_{7/2}}& 4277.8&1.000& 0.15&--- \\ 
\mult{3d\ ^4P_{5/2}}{4f\ D[2]^o_{5/2}}& 4281.3&1.000& 0.08& 0.19   \\ 
\mult{3d\ ^4D_{3/2}}{4f\ F[2]^o_{5/2}}& 4282.9&1.000& 0.24& 0.26   \\ 
\mult{3d\ ^4D_{3/2}}{4f\ F[2]^o_{3/2}}& 4283.6&1.000& 0.15&--- \\ 
\mult{3d\ ^2F_{5/2}}{4f\ F[3]^o_{7/2}}& 4285.6&1.000& 0.29& 0.31   \\ 
\mult{3d\ ^4P_{5/2}}{4f\ G[3]^o_{7/2}}& 4291.2&1.000& 0.24&--- \\ 
\hline

\end{tabular}

{}*\lam4097.2 includes emission of \mult{3p\ ^4P^o_{1/2}}{3d\ ^4D_{3/2}}\\
$\dagger$ 3p' and 3d' are core excited configurations \term{2p^2(^1D)3p} and 
\term{2p^2(^1D)3d} respectively.
\end{table}
\begin{table}
\contcaption{}
\begin{tabular}{llclll}
\hline
Lower--Upper&\lam(\AA)&$I_{\rm rec}/I_{\rm calc}$&$I_{\rm calc}$&$I_{\rm obs}$\\
\hline
\mult{3d\ ^2F_{5/2}}{4f\ F[2]^o_{5/2}}& 4292.2&1.000& 0.14&--- \\ 
\mult{3d\ ^4P_{3/2}}{4f\ D[2]^o_{5/2}}& 4294.8&1.000& 0.36& 0.59   \\ 
\mult{3d\ ^4D_{7/2}}{4f\ G[5]^o_{9/2}}& 4303.5&1.000& 0.09&--- \\ 
\mult{3d\ ^4P_{5/2}}{4f\ D[3]^o_{7/2}}& 4303.8&1.000& 0.63& 0.66   \\ 
\mult{3d\ ^4P_{1/2}}{4f\ D[2]^o_{3/2}}& 4307.3&1.000& 0.16& 1.19   \\ 
\mult{3d\ ^2F_{7/2}}{4f\ F[4]^o_{7/2}}& 4312.2&1.000& 0.10&--- \\ 
\mult{3d\ ^2F_{7/2}}{4f\ F[4]^o_{9/2}}& 4313.5&1.000& 0.19&--- \\ 
\mult{3s\ ^4P_{1/2}}{3p\ ^4P^o_{3/2}}& 4317.0&0.605& 1.58& 1.59   \\ 
\mult{3d\ ^4P_{3/2}}{4f\ D[3]^o_{5/2}}& 4317.7&1.000& 0.11&--- \\ 
\mult{3s\ ^4P_{3/2}}{3p\ ^4P^o_{5/2}}& 4319.6&0.747& 1.25& 0.83 ? \\ 
\mult{3s\ ^4P_{1/2}}{3p\ ^4P^o_{1/2}}& 4325.8&0.522& 0.34&--- \\ 
\mult{3d\ ^4D_{5/2}}{4f\ G[4]^o_{7/2}}& 4331.2&1.000& 0.12&--- \\ 
\mult{3d\ ^4D_{7/2}}{4f\ G[4]^o_{9/2}}& 4332.6&1.000& 0.15&--- \\ 
\mult{3s\ ^4P_{3/2}}{3p\ ^4P^o_{3/2}}& 4336.7&0.605& 0.67& 0.37   \\ 
\mult{3d\ ^2F_{5/2}}{4f\ G[4]^o_{7/2}}& 4340.3&1.000& 0.34&--- \\ 
\mult{3d\ ^2F_{7/2}}{4f\ G[5]^o_{9/2}}& 4342.1&1.000& 0.85&--- \\ 
\mult{3d\ ^4D_{5/2}}{4f\ G[3]^o_{7/2}}& 4344.4&1.000& 0.18&--- \\ 
\mult{3s\ ^4P_{3/2}}{3p\ ^4P^o_{1/2}}& 4345.6&0.522& 1.92& 1.92   \\ 
\mult{3s'\ ^2D_{3/2}}{3p'\ ^2D^o_{3/2}}$\dagger$& 4347.5&0.649& 0.57&--- \\ 
\mult{3s\ ^4P_{5/2}}{3p\ ^4P^o_{5/2}}& 4349.5&0.747& 3.51& 3.30   \\ 
\mult{3s'\ ^2D_{5/2}}{3p'\ ^2D^o_{5/2}}& 4351.2&0.571& 0.69& 0.80   \\ 
\mult{3d\ ^2F_{5/2}}{4f\ G[3]^o_{7/2}}& 4353.5&1.000& 0.16&--- \\ 
\mult{3s\ ^4P_{5/2}}{3p\ ^4P^o_{3/2}}& 4366.8&0.605& 1.67& 1.36   \\ 
\mult{3d\ ^2F_{7/2}}{4f\ G[4]^o_{9/2}}& 4371.7&1.000& 0.15&--- \\ 
\mult{3s\ ^2P_{3/2}}{3p\ ^2D^o_{5/2}}& 4414.9&0.812& 5.26& 3.03   \\ 
\mult{3s\ ^2P_{1/2}}{3p\ ^2D^o_{3/2}}& 4416.8&0.815& 2.95& 1.97   \\ 
\mult{3s\ ^2P_{3/2}}{3p\ ^2D^o_{3/2}}& 4452.2&0.815& 0.54& 0.88 ? \\ 
\mult{3d\ ^2P_{3/2}}{4f\ D[2]^o_{5/2}}& 4466.4&1.000& 0.14& 0.29   \\ 
\mult{3d\ ^2P_{3/2}}{4f\ G[3]^o_{5/2}}& 4477.9&1.000& 0.13&--- \\ 
\mult{3d\ ^2P_{1/2}}{4f\ D[2]^o_{3/2}}& 4489.4&1.000& 0.10& 0.07   \\ 
\mult{3d\ ^2P_{3/2}}{4f\ D[3]^o_{5/2}}& 4491.2&1.000& 0.21&--- \\ 
\mult{3s'\ ^2D_{5/2}}{3p'\ ^2F^o_{7/2}}& 4590.9&0.729& 0.60& 1.43   \\ 
\mult{3s'\ ^2D_{3/2}}{3p'\ ^2F^o_{5/2}}& 4596.2&0.774& 0.53& 0.94   \\ 
\mult{3d\ ^2D_{3/2}}{4f\ F[3]^o_{5/2}}& 4602.1&1.000& 0.27& 0.22   \\ 
\mult{3d\ ^2D_{5/2}}{4f\ F[4]^o_{7/2}}& 4609.3&1.000& 0.66& 0.60   \\ 
\mult{3d\ ^2D_{3/2}}{4f\ F[2]^o_{5/2}}& 4610.2&1.000& 0.21&--- \\ 
\mult{3s\ ^4P_{1/2}}{3p\ ^4D^o_{3/2}}& 4638.8&0.731& 2.07& 1.96   \\ 
\mult{3s\ ^4P_{3/2}}{3p\ ^4D^o_{5/2}}& 4641.6&0.794& 4.68& 4.22   \\ 
\mult{3s\ ^4P_{5/2}}{3p\ ^4D^o_{7/2}}& 4649.2&0.795& 8.35& 6.70   \\ 
\mult{3s\ ^4P_{1/2}}{3p\ ^4D^o_{1/2}}& 4650.9&0.688& 2.05& 2.17   \\ 
\mult{3s\ ^4P_{3/2}}{3p\ ^4D^o_{3/2}}& 4661.6&0.731& 2.30& 2.31   \\ 
\mult{3s\ ^4P_{3/2}}{3p\ ^4D^o_{1/2}}& 4673.8&0.688& 0.35& 0.41   \\ 
\mult{3s\ ^4P_{5/2}}{3p\ ^4D^o_{5/2}}& 4676.2&0.794& 1.59& 1.58   \\ 
\mult{3s\ ^4P_{5/2}}{3p\ ^4D^o_{3/2}}& 4696.4&0.731& 0.17& 0.27   \\ 
\mult{3p\ ^2D^o_{3/2}}{3d\ ^2F_{5/2}}& 4699.3&0.441& 0.70& 1.33 ? \\ 
\mult{3p\ ^2D^o_{5/2}}{3d\ ^2F_{7/2}}& 4705.2&0.472& 1.36& 1.83   \\ 
\mult{3p\ ^4S^o_{3/2}}{3d\ ^4D_{5/2}}& 4856.4&0.686& 0.11&--- \\ 
\mult{3p\ ^4S^o_{3/2}}{3d\ ^4P_{1/2}}& 4890.8&0.465& 0.41& 1.37   \\ 
\mult{3p\ ^4S^o_{3/2}}{3d\ ^4P_{3/2}}& 4906.9&0.655& 0.71& 0.42   \\ 
\mult{3p\ ^4S^o_{3/2}}{3d\ ^4P_{5/2}}& 4924.7&0.711& 0.41& 0.89   \\ 
\hline
\multicolumn{2}{r}{$\langle I_{\rm rec}/I_{\rm calc}\rangle=0.838$}
&\multicolumn{3}{r}{$\langle I_{\rm calc}/I_{\rm obs}\rangle=0.898$}\\
\hline

\end{tabular}

$\dagger$ 3p' and 3d' are core excited terms \term{2p^2(^1D)3p} and 
\term{2p^2(^1D)3d} respectively.
\end{table}
\begin{table}
\caption{Comparison of predicted and observed line 
intensities of \oii\ with $A$--values from (1) \citet{nahar10} 
and (2) NIST.}
\label{avalues}
\tabcolsep=3pt
\begin{tabular}{lllllll}
\hline
Lower--Upper&\lam(\AA)&$I_{\rm obs}$
&\multicolumn{2}{c}{$A\,(10^6\persecond)$}
&\multicolumn{2}{c}{$I_{\rm calc}/I_{\rm obs}$}\\
&&&(1)&(2)&(1)&(2)\\
\hline
\mult{3p\ ^4D^o_{5/2}}{3d\ ^4P_{5/2}}&3907.6&0.35&1.55&8.64&0.059&0.254\\
\mult{3p\ ^4P^o_{3/2}}{3d\ ^4D_{1/2}}&4110.9&0.75&0.0321&80.5&0.00004&0.090\\
\mult{3p\ ^4P^o_{1/2}}{3d\ ^4P_{1/2}}&4121.3&1.66&6.82&56.0&0.038&0.245\\
\mult{3p\ ^4P^o_{3/2}}{3d\ ^4P_{5/2}}&4153.4&1.84&1.09&72.8&0.007&0.382\\
\hline
\end{tabular}
\end{table}

\subsection{The velocity field}
\label{velocity}

The pumping transitions in fluorescence excitation 
will be Doppler shifted in an expanding nebula, 
and absorb the stellar flux at longer wavelengths.  
Our high--dispersion SED gives the chance 
to notice the effect of the velocity field of 
the gas on the intensities of some lines mostly excited by 
fluorescence. 
Several authors have used different velocity 
fields to model forbidden--line profiles 
in PN \citep{neiner, gesicki, zhang}. 
We have assumed two accelerating velocity fields, 
$V\propto R$, and $V\propto R^4$ with a maximum 
velocity of 40\kms at the outer edge of the nebula, 
which are similar to fields considered in the 
literature. 
Tables~\ref{tabcii} to~\ref{taboii} were computed 
with the $V\propto R^4$ field.  
Although PN shells are expected to expand, 
we also considered a static field with zero 
velocity for comparison purposes. 

Turbulence is another component of the velocity 
field that must be considered. 
Turbulence of 14\kms\ was needed 
to model broad line profiles in a PN with a Wolf--Rayet 
central star, 
but was not needed for O--star PNe by \citet{neiner}. 
\citet{chrisgrasi}, however, have shown that 
broadened line profiles in PNe can be explained with 
a non--spherical geometry with negligible turbulence. 
In our case turbulence increases the pumping probability 
because it increases the width of the 
Voigt profile of a pumping transition 
(see equations~[3] and [4] of EM). 
We considered a moderate turbulent field of 3\kms. 
Table~\ref{tabvels} shows some of the lines that exhibit the largest 
variations in intensity with different velocity fields. 

The effect of a non--zero expansion velocity is 
stronger in lines that are pumped 
by a single transition from the ground or a metastable 
state like the \nii\ lines from the states  
\term{3d ^3P^o_0}, \term{3d\ ^1D^o_2} 
and \term{3d\ ^1F^o_3}. 
The intensities of those lines generally change 
by amounts larger than the assumed observational uncertainty 
between 20 and 80 per cent in the velocity field that we have 
adopted with respect to a static gas, 
and tend to match the observations better than 
with a static field. 
The \nii\ \mult{3d\ ^1P_1}{3d\ ^1D^o_2}\lam4447.0 line 
increases by a factor of 2 with respect to a static field 
because the line is mostly pumped by the transitions   
\mult{2p\ ^3P_1}{3d\ ^1D^o_2}\lam534.657 and
\mult{2p^2\ ^1D_2}{3d\ ^1D^o_3}\lam582.156, 
which lie next to peaks of the SED at longer 
wavelengths (see Fig.~\ref{sed}). 

\begin{table*}
\centering
\caption{Variation of normalized intensity of selected lines with 
velocity field. 
(1) $V=0$ and no turbulence, 
(2) $V\propto R$ and no turbulence, 
(3) $V\propto R^4$ and no turbulence, 
(4) $V\propto R$ and 3\kms turbulence, 
(5) $V\propto R^4$ and 3\kms turbulence. 
}
\label{tabvels}
  \begin{tabular}{llcccccl}
  \hline
Lower--Upper&\lam(\AA)&\multicolumn{5}{c}{$I_{\rm calc}/I_{\rm obs}$}&
$I_{\rm obs}$\\
&&(1)&(2)&(3)&(4)&(5)\\
  \hline
         \multicolumn{8}{c}\cii\\
  \hline
\mult{3p\ ^2P^o_{1/2}}{4s\ ^2S_{1/2}}&3919.0&0.64& 0.64& 0.63& 0.85& 0.83&10.68\\
\mult{3p\ ^2P^o_{3/2}}{4s\ ^2S_{1/2}}&3920.7&0.66& 0.66& 0.65& 0.88& 0.86&20.52\\
\mult{4p\ ^2P^o_{3/2}}{3p'\ ^2P_{1/2}}&5127.0&0.87&0.88&0.87&1.16& 1.14& 0.30 ?\\
\mult{2p^3\ ^2P^o_{1/2}}{3p'\ ^2P_{1/2}}&7519.9&0.73&0.91&0.73&0.98&0.97&0.92 ?\\
\mult{2p^3\ ^2P^o_{3/2}}{3p'\ ^2P_{1/2}}&7530.6&0.88&0.90&0.89&1.18&1.16&0.38 ?\\
\mult{4p\ ^2P^o_{3/2}}{3p'\ ^2P_{3/2}}&5121.8& 0.28&0.27&0.28&0.34&0.35& 2.68 ?\\
\mult{2p^3\ ^2P^o_{3/2}}{3p'\ ^2P_{3/2}}&7519.5&0.93&0.74&0.73&1.16&1.20&1.03 ?\\
\mult{4p\ ^2P^o_{3/2}}{3p'\ ^2D_{5/2}}&3831.7& 0.53&0.78&0.62&0.94&0.70& 3.00 ?\\
\mult{2p^3\ ^2P^o_{3/2}}{3p'\ ^2D_{5/2}}&5032.1&0.43&0.64&0.50&0.76&0.57&4.34 ?\\
\mult{4p\ ^2P^o_{1/2}}{5d\ ^2D_{3/2}}&6257.2& 1.14&0.94& 1.10& 1.08& 1.29& 0.52\\
\mult{4p\ ^2P^o_{3/2}}{5d\ ^2D_{5/2}}&6259.6& 0.70&0.60& 0.68& 0.65& 0.76& 0.84\\
  \hline
         \multicolumn{8}{c}\nii\\
  \hline
\mult{3p\ ^3D_1}{3d\ ^3P^o_0}& 4459.9& 0.43& 0.23& 0.32& 0.25& 0.35& 0.25   \\
\mult{3p\ ^3D_2}{3d\ ^3P^o_1}& 4477.7& 0.48& 0.61& 0.62& 0.68& 0.70& 0.61   \\
\mult{3p\ ^3S_1}{3d\ ^3P^o_1}& 4994.4& 0.54& 0.68& 0.68& 0.76& 0.78& 4.23   \\
\mult{3p\ ^3P_2}{3d\ ^3P^o_2}& 5495.7& 0.46& 0.34& 0.41& 0.38& 0.48& 1.48   \\
\mult{3p\ ^3D_1}{3d\ ^3D^o_1}& 4779.7& 0.66& 0.47& 0.62& 0.51& 0.68& 1.79   \\
\mult{3p\ ^3P_0}{3d\ ^3D^o_1}& 5927.8& 0.63& 0.45& 0.60& 0.50& 0.65& 1.91   \\
\mult{3p\ ^1P_1}{3d\ ^1D^o_2}& 4447.0& 0.68& 1.40& 1.12& 1.37& 1.11& 0.39   \\
\mult{3p\ ^1D_2}{3d\ ^1F^o_3}& 6610.6& 0.50& 0.78& 0.62& 0.77& 0.61& 0.28   \\
  \hline
         \multicolumn{8}{c}\oii\\
  \hline
\mult{3p\ ^4P^o_{1/2}}{3d\ ^4P_{1/2}}&4121.3&0.23& 0.35& 0.23& 0.38& 0.24& 1.66\\
\mult{3p\ ^4S^o_{3/2}}{3d\ ^4P_{1/2}}&4890.8&0.28& 0.43& 0.28& 0.46& 0.30& 1.37\\
\mult{3p\ ^4D^o_{5/2}}{3d\ ^4P_{5/2}}&3907.6&0.25& 0.35& 0.25& 0.34& 0.25& 0.35\\
\mult{3p\ ^4P^o_{3/2}}{3d\ ^4P_{5/2}}&4153.4&0.38& 0.52& 0.38& 0.52& 0.38& 1.84\\
\mult{3p\ ^4S^o_{3/2}}{3d\ ^4P_{5/2}}&4924.7&0.46& 0.63& 0.46& 0.62& 0.46& 0.89\\
\mult{3p\ ^4P^o_{3/2}}{3d\ ^4D_{3/2}}&4105.1&1.47& 1.40& 1.44& 1.59& 1.65& 0.46\\
\mult{3p\ ^2D^o_{3/2}}{3d\ ^2F_{5/2}}&4699.3&0.57&0.60&0.54& 0.60& 0.52& 1.33 ?\\
\mult{3p\ ^2D^o_{5/2}}{3d\ ^2F_{7/2}}&4705.2&0.64& 0.97& 0.77& 0.96& 0.74& 1.83\\
  \hline
\end{tabular}

\end{table*}

Lines that are pumped by more than one transition 
generally do not show large changes because variations 
in the intensity of the stellar flux at the wavelengths 
in the rest frame of the absorber tend to cancel out 
for the different transitions in a multiplet. 
One exception are lines from the 
\nii\ \term{3d ^3P^o_1} state with two main pumping 
transitions from the ground term at 
\mult{2p^2\ ^3P_0}{3d\ ^3P^o_1}\lam529.355 
and \mult{2p^2\ ^3P_1}{3d\ ^3P^o_1}\lam529.491,  
which receive an increased stellar flux simultaneously at 
a redshifted wavelength. 

The sensitivity of lines affected by fluorescence on the 
velocity gradient depends on the ion spatial distribution 
within the nebula. 
The variation of intensities between the static and 
accelerating fields with no turbulence is comparable or 
slightly larger than our assumed uncertainty of 20 per cent 
for most of the \nii\ and \oii\ lines in table~\ref{tabvels}. 
Since the \op\ ion is more extended towards the inner parts of 
the nebula, 
\oii\ lines are more sensitive to the field gradient, 
with variations in intensity larger than the assumed uncertainty 
in the $V\propto R$ field. 
The \cp\ and \np\ ions are more concentrated towards the outer parts. 
Their fluorescence lines are nearly insensitive to the choice 
of velocity field with the exception of the three \nii\ lines 
\mult{3p\ ^3D_1}{3d\ ^3P^o_0}\lam4459.94, 
\mult{3p\ ^1P_1}{3d\ ^1D^o_2}\lam4447.03 and 
\mult{3p\ ^1D_2}{3d\ ^1F^o_3}\lam6610.56 from the states mentioned 
above. 

Turbulence is the most important factor in the intensity 
of the \cii\ lines because of the high \cp\ column density and 
optical depth of its pumping transitions. Most \cii\ lines 
affected by fluorescence increase their intensity by 30 per cent in 
a 3\kms\ turbulent field with respect to a field without turbulence. 

\subsection{\ni} 

Fluorescence is important in atomic nitrogen because 
several resonant lines are just below the Lyman limit. 
The observed \ni\ dipole--allowed lines are quartets in the 
3s--3p array mainly produced by 
absorptions in multiplets \mult{2p^3\ ^4S^o}{3d\ ^4P}
\lala953.415--953.970 and 
\mult{2p^3\ ^4S^o}{4s\ ^4P}\lala963.041--964.626 
followed by decays to the 3p terms. 
SBW noted that the similarity of 
the line profiles of quartets and the \ni\ forbidden lines 
points to fluorescence as the main excitation mechanism of 
the \ni\ quartets. 
A 3\kms\ turbulence increases the intensity of the 
\ni\ quartets by a factor of 3 because of the large 
optical depth of the pumping transitions, 
and brings them within the uncertainty interval of the 
observed intensities. 

Fluorescence contributes 50 per cent of the forbidden 
line \mult{2p^3\ ^4S^o_{3/2}}{^2D^o_{5/2}}\lam5200.26 
and most of the forbidden line 
\mult{2p^3\ ^4S^o_{3/2}}{^2D^o_{3/2}}\lam5197.90 
due to absorptions from the metastable term \term{2p^3\ ^2D^o}
and to the unusually intense spin--forbidden transition 
\mult{2p^3\ ^4S^o_{3/2}}{3d\ ^2F_{5/2}}\lam954.104. The 
state \term{3d\ ^2F_{3/2}} preferentially decays 
to state \term{2p^3\ ^2D^o_{3/2}} besides other doublets 
producing half of the total fluorescence contribution to the 
forbidden lines. If fluorescence is turned off, however, 
electron collisions alone can substitute the role of 
fluorescence in the excitation of the forbidden lines 
\lam5200.26 and \lam5197.90. 
The reason is that the population of the ground state 
increases with the lack of absorptions to excited states, 
and thus the rate of collisional excitations also increases. 
The other forbidden line, 
\mult{2p^3\ ^4S^o_{3/2}}{^2P^o_{1/2,3/2}}\lala3466.50, 3466.54, 
is more heavily influenced by fluorescence, but it is 
beyond the red end of the spectrum of SWBH. 

The fluorescence contribution to the forbidden lines 
depends on the A--value of the spin--forbidden \lam954.104 
mentioned above. Intermediate calculations by \citet{hibbert} 
gave a value of $7.609\times10^5\persecond$, 
but the calculation by \citet{charlotte} 
gave $1.954\times10^7\persecond$ and \citet{goldbach} 
measured $3.30\times10^7\persecond$, which is the value 
listed in the NIST database and adopted by us. 

The measured intensities of the forbidden lines show the 
largest discrepancy with our predicted values for \ni. 
The agreement with the fluorescence lines improves 
if the density and size of the PDR are increased, 
but the forbidden lines remain underestimated for 
reasonable values of the PDR parameters in \ecua{pdr}. 
As shown by EM in Orion, 
the efficiency of fluorescence excitation grows 
in a manner analogous of the curve of growth. 
An increasing column density eventually saturates 
the pumping transitions and the fluorescence 
excitation grows logarithmically in the line wings. 

We only added the two lines 
\mult{3s\ ^4P_{1/2}}{3p\ ^4P^o_{3/2}}\lam8188.01  
and \mult{3s\ ^4P_{3/2}}{3p\ ^4P^o_{3/2}}\lam8210.72  
from SWBH to the selection in SBW. 
The multiplet \mult{3p\ ^4D^o}{3d\ ^4D}\lala9776.90--9872.15 has 
three intense lines detected by SWBH. 
Its other detectable components at longer wavelengths 
are beyond the observed spectral range. 

\begin{table}
\caption{Predicted and observed line intensities of \ni.}
\label{tabni}
\begin{tabular}{lrclll}
\hline
Lower--Upper&\lam(\AA)&$I_{\rm rec}/I_{\rm calc}$&$I_{\rm calc}$&$I_{\rm obs}$\\
\hline
\mult{2p^3\ ^4S^o_{3/2}}{2p^3\ ^2D^o_{3/2}}& 5197.9&0.000& 3.58&20.11   \\ 
\mult{2p^3\ ^4S^o_{3/2}}{2p^3\ ^2D^o_{5/2}}& 5200.3&0.000& 2.21&11.73   \\ 
\mult{3s\ ^4P_{1/2}}{3p\ ^4S^o_{3/2}}& 7423.6&0.007& 1.62& 1.56   \\ 
\mult{3s\ ^4P_{3/2}}{3p\ ^4S^o_{3/2}}& 7442.3&0.007& 3.36& 3.63   \\ 
\mult{3s\ ^4P_{5/2}}{3p\ ^4S^o_{3/2}}& 7468.3&0.007& 5.21& 5.83   \\ 
\mult{3s\ ^4P_{3/2}}{3p\ ^4P^o_{5/2}}& 8184.9&0.017& 1.81& 1.52   \\ 
\mult{3s\ ^4P_{1/2}}{3p\ ^4P^o_{3/2}}& 8188.0&0.007& 4.36& 3.81 ? \\ 
\mult{3s\ ^4P_{1/2}}{3p\ ^4P^o_{1/2}}& 8200.4&0.005& 1.20& 1.31   \\ 
\mult{3s\ ^4P_{3/2}}{3p\ ^4P^o_{3/2}}& 8210.7&0.007& 1.66& 1.98 ? \\ 
\mult{3s\ ^4P_{5/2}}{3p\ ^4P^o_{5/2}}& 8216.3&0.017& 4.69& 5.98   \\ 
\mult{3s\ ^4P_{3/2}}{3p\ ^4P^o_{1/2}}& 8223.1&0.005& 6.37& 6.23   \\ 
\mult{3s\ ^4P_{5/2}}{3p\ ^4P^o_{3/2}}& 8242.4&0.007& 4.64& 5.51   \\ 
\mult{3s\ ^4P_{5/2}}{3p\ ^4D^o_{7/2}}& 8680.3&0.041& 3.40& 3.85   \\ 
\mult{3s\ ^4P_{3/2}}{3p\ ^4D^o_{5/2}}& 8683.4&0.022& 3.50& 4.17   \\ 
\mult{3s\ ^4P_{1/2}}{3p\ ^4D^o_{3/2}}& 8686.1&0.015& 2.12& 2.20   \\ 
\mult{3s\ ^4P_{1/2}}{3p\ ^4D^o_{1/2}}& 8703.2&0.013& 2.36& 2.68   \\ 
\mult{3s\ ^4P_{3/2}}{3p\ ^4D^o_{3/2}}& 8711.7&0.015& 2.48& 2.73   \\ 
\mult{3s\ ^4P_{5/2}}{3p\ ^4D^o_{5/2}}& 8718.8&0.022& 1.31& 1.35   \\ 
\mult{3s\ ^4P_{3/2}}{3p\ ^4D^o_{1/2}}& 8728.9&0.013& 0.42&--- \\ 
\mult{3s\ ^4P_{5/2}}{3p\ ^4D^o_{3/2}}& 8747.4&0.015& 0.20&--- \\ 
\mult{3p\ ^4D^o_{1/2}}{3d\ ^4D_{3/2}}& 9776.9&0.003& 0.19&--- \\ 
\mult{3p\ ^4D^o_{3/2}}{3d\ ^4D_{5/2}}& 9786.8&0.005& 0.17&--- \\ 
\mult{3p\ ^4D^o_{1/2}}{3d\ ^4D_{1/2}}& 9788.3&0.002& 0.42&--- \\ 
\mult{3p\ ^4D^o_{3/2}}{3d\ ^4D_{3/2}}& 9798.6&0.003& 0.44&--- \\ 
\mult{3p\ ^4D^o_{3/2}}{3d\ ^4D_{1/2}}& 9810.0&0.002& 0.74& 0.82 ? \\ 
\mult{3p\ ^4D^o_{5/2}}{3d\ ^4D_{5/2}}& 9822.7&0.005& 0.76& 0.76 ? \\ 
\mult{3p\ ^4D^o_{5/2}}{3d\ ^4D_{3/2}}& 9834.6&0.003& 0.72& 0.82 ? \\ 
\mult{3p\ ^4D^o_{7/2}}{3d\ ^4D_{5/2}}& 9872.1&0.005& 0.45&--- \\ 
\mult{3p\ ^4P^o_{1/2}}{3d\ ^4D_{3/2}}&10500.3&0.003& 0.92&--- \\ 
\mult{3p\ ^4P^o_{3/2}}{3d\ ^4D_{5/2}}&10507.0&0.005& 1.79&--- \\ 
\mult{3p\ ^4P^o_{1/2}}{3d\ ^4D_{1/2}}&10513.4&0.002& 2.37&--- \\ 
\mult{3p\ ^4P^o_{3/2}}{3d\ ^4D_{3/2}}&10520.6&0.003& 2.29&--- \\ 
\mult{3p\ ^4P^o_{3/2}}{3d\ ^4D_{1/2}}&10533.8&0.002& 0.95&--- \\ 
\mult{3p\ ^4P^o_{5/2}}{3d\ ^4D_{5/2}}&10549.6&0.005& 1.64&--- \\ 
\mult{3p\ ^4P^o_{5/2}}{3d\ ^4D_{3/2}}&10563.3&0.003& 0.49&--- \\ 
\hline
\multicolumn{2}{r}{$\langle I_{\rm rec}/I_{\rm calc}\rangle=0.010$}
&\multicolumn{3}{r}{$\langle I_{\rm calc}/I_{\rm obs}\rangle=0.865$}\\
\hline

\end{tabular}

\end{table}

\begin{table}
\caption{Predicted and observed line intensities of \oi.}
\label{taboi}
\begin{tabular}{llclll}
\hline
Lower--Upper&\lam(\AA)&$I_{\rm rec}/I_{\rm calc}$&$I_{\rm calc}$&$I_{\rm obs}$\\
\hline
\mult{3s\ ^3S^o_1}{4p\ ^3P} \S& 4368.2&0.011&12.30& 9.64 ? \\ 
\mult{3p\ ^3P}{7d\ ^3D^o}& 5275.1&0.000& 6.34& 1.86 ? \\ 
\mult{3p\ ^3P}{8s\ ^3S^o_1}& 5298.9&0.000& 5.01& 3.71 ? \\ 
\mult{3p\ ^5P_3}{5d\ ^5D^o_2}& 5330.7&1.000& 0.14& 0.23 ? \\ 
\mult{3p\ ^3P}{6d\ ^3D^o}& 5512.8&0.001& 8.64& 3.13 ? \\ 
\mult{3p\ ^3P}{7s\ ^3S^o_1}& 5554.8&0.000& 6.87& 1.59 ? \\ 
\mult{2p^4\ ^1D_2}{2p^4\ ^1S_0}& 5577.3&0.000& 1.64& 2.63   \\ 
\mult{3p\ ^3P}{5d\ ^3D^o}& 5958.6&0.003& 5.91& 4.75 ? \\ 
\mult{3p\ ^3P}{6s\ ^3S^o_1}& 6046.2&0.000&11.20& 8.81 ? \\ 
\mult{3p\ ^5P_1}{4d\ ^5D^o}& 6156.0&1.000& 0.10& 0.18 ? \\ 
\mult{3p\ ^5P_2}{4d\ ^5D^o}& 6156.7&1.000& 0.17& 0.36 ? \\ 
\mult{3p\ ^5P_3}{4d\ ^5D^o}& 6158.1&1.000& 0.23& 0.42 ? \\ 
\mult{2p^4\ ^3P_2}{2p^4\ ^1D_2}& 6300.3&0.000&146.00&217.53   \\ 
\mult{2p^4\ ^3P_1}{2p^4\ ^1D_2}& 6363.8&0.000&46.80&75.94   \\ 
\mult{3p\ ^5P_2}{5s\ ^5S^o_2}& 6454.4&1.000& 0.07& 0.10 ? \\ 
\mult{3p\ ^5P_3}{5s\ ^5S^o_2}& 6456.0&1.000& 0.10& 0.13 ? \\ 
\mult{3p\ ^3P}{4d\ ^3D^o}& 7001.9&0.005&10.80& 8.49 ? \\ 
\mult{3p\ ^3P}{5s\ ^3S^o_1}& 7254.2&0.002& 8.57&15.64 ? \\ 
\mult{3s\ ^5S^o_2}{3p\ ^5P_3}& 7771.9&1.000& 1.95& 3.52   \\ 
\mult{3s\ ^5S^o_2}{3p\ ^5P_2}& 7774.2&0.999& 1.39& 2.15   \\ 
\mult{3s\ ^5S^o_2}{3p\ ^5P_1}& 7775.4&1.000& 0.84& 1.30   \\ 
\mult{3s\ ^3S^o_1}{3p\ ^3P}& 8446.4&0.017&135.00&114.19   \\ 
\mult{3p\ ^5P_1}{3d\ ^5D^o}& 9260.8&1.000& 0.31& 0.47   \\ 
\mult{3p\ ^5P_2}{3d\ ^5D^o}& 9262.6&1.000& 0.52& 0.97   \\ 
\mult{3p\ ^5P_3}{3d\ ^5D^o}& 9265.8&1.000& 0.73& 1.25   \\ 
\hline
\multicolumn{2}{r}{$\langle I_{\rm rec}/I_{\rm calc}\rangle=0.460$}
&\multicolumn{3}{r}{$\langle I_{\rm calc}/I_{\rm obs}\rangle=1.096$}\\
\hline

\end{tabular}

\S\ Terms with no subscript are blended transitions of several states 
of angular momentum $J$.\\
\end{table}

\subsection{\oi}
\label{sectoi}

The triplet terms are excited by fluorescence in the PDR of \ic\ 
through a large number of resonant transitions of comparable 
strength from the ground term \term{2p^2\ ^3P}, 
a situation also encountered in Orion by \citet{lucy}. 
Triplet \oi\ line profiles shown by SBW are similar to those 
of the \oi\ forbidden lines in support of the fluorescence 
excitation mechanism. 
The quintets on the other hand 
are produced in the ionized region by recombination of 
\op\ with no contribution from fluorescence because they are 
not connected with strong transitions from the ground term and 
recombination preferentially populates higher multiplicity states. 

SWBH identified several multiplets between terms \term{3p\ ^3P} 
and \term{3p\ ^5P} and s states with principal quantum number 
$n=5$ to 8 and d states with principal quantum number 
$n=4$ to 7. SBW does not list them although their identifications 
seem certain and only their effective recombinations are 
lacking in \citet{pequignot}. 
The effective recombination coefficients for those terms 
were calculated by \citet{escvict92}. 
Our calculations tend to underestimate systematically the quintet 
and some triplet intensities by a factor of 1.3, 
which is slightly higher than the assumed uncertainty
and suggests that the effective recombination coefficients 
of \citet{escvict92} may be underestimated. 
The effective recombination coefficients of \citet{pequignot} 
are higher by about that same factor, and thus give a 
closer agreement with the observations. 

The forbidden lines of the ground configuration 
\mult{2p^4\ ^3P_2}{2p^4\ ^1D_2}\lam6300.30, 
\mult{2p^4\ ^3P_1}{2p^4\ ^1D_2}\lam6363.78 and 
\mult{2p^4\ ^1D_2}{2p^4\ ^1S_1}\lam5577.34 
are excited by electron collisions with negligible 
contribution from fluorescence in the warm interface of 
the PDR and ionized region. 
Our predicted intensities for those lines are 
underestimated by the same amount as the recombination lines, 
but they are more sensitive to the density 
than the \ni\ optical forbidden lines. 
If the $n_3$ parameter in \ecua{pdr} 
is raised to 35000, our predicted intensities 
of the three forbidden lines perfectly match the 
observed values. 
This value of the density is already too high 
to explain the FIR [\cii] and [\oi] line 
observations of \citet{liu01a}. 
It is also possible to improve the agreement 
of the calculated fluorescence and recombination 
lines with the measurements if the PDR were larger, 
but there is no evidence of a dense and extended 
neutral medium around \ic. 
Such a medium would increase the FIR line fluxes beyond 
their observed uncertainty. 

The largest discrepancy with the observations 
occurs in multiplet 
\mult{3p\ ^3P_{0,1,2}}{5s\ ^3S^o_1}\lam7254.15--.53  
with an observed intensity 1.8 times the predicted one. 
The possibility of an overestimated observed intensity due 
to a low S/N \citep{rola} can be ruled out because 
the S/N measured by SWBH for the \lam7254 multiplet is 15.0. 
There is only one \oi\ line with $\rm S/N<7$ in our calculations. 
Most of the \oi\ lines or blends are relatively intense with $\rm S/N>100$. 
Lines from other states in the Rydberg series, \term{6s\ ^3S^o_1}, 
and \term{8s\ ^3S^o_1} agree better with the calculated values. 

SWBH also identified the feature at \lam4654.475 as 
the \oi\ multiplet \mult{3p\ ^5P_2}{8d\ ^5D^o_{1,2,3}}\la4654.56,  
which is strongly underestimated by a factor of 
11 in our calculations and is not listed in table~\ref{taboi}. 
We notice that the other terms in the series 
\term{6d\ ^5D} and \term {7d\ ^5D} were not 
detected by SWBH. 
For example, the other multiplets in the Rydberg series, 
\mult{3p\ ^5P_2}{6d\ ^5D^o_{1,2,3}} at \lam4967.88 
and \mult{3p\ ^5P_2}{7d\ ^5D^o_{1,2,3}} at \lam4772.91 
should be theoretically stronger by 2.3 and 1.4 times the 
intensity of the \lam4654.56 multiplet respectively 
according to the recombination rates of \citet{escvict92}, 
and should be above the noise in their respective parts of 
the spectrum. 
As noted in section~\ref{sectnii}, an identification as a 
\nii\ line is more likely, 
and therefore the line at \lam4654.56 is probably a misidentification 
as a \oi\ line.

\subsection{Variation of the SED}

We compared results calculated with the original SED of MG 
and the new SED described in section~\ref{stellar}  
to gauge the dependence of the 
fluorescence excitation on theoretical assumptions about 
the stellar atmosphere. 

With few exceptions, most of the variation in results occurs in 
the \nii\ lines because of their high fluorescence excitation. 
The new SED increases the intensities of some \nii\ lines
from p states by 25 per cent and the lines originating in the 
\term{3d\ ^3D^o_2} state by a factor of 2. 
The only significant decrease is a factor of 2 in the line 
\mult{3p\ ^3D_1}{3d\ ^3P^o_0}\lam4459.937. 
We traced back the cause of those variations to an increase 
of the stellar flux in the new SED at the wavelength of the 
transition \mult{2p^2\ ^3P_2}{3d\ ^3D^o_2}\lam533.815, 
and a decrease at the wavelength of the 
transition \mult{2p^2\ ^3P_1}{3d\ ^3P^o_0}\lam529.413,  
which pump the observed lines. 
Minor variations occur in \nii\ singlet lines 
\mult{3p^1P_1}{3d^1D^o_2}\lam4447.03 and 
\mult{3p^1D_2}{3d^1F^o_3}\lam6610.56,  
which are strongly excited by fluorescence. 

\section{Summary}

We have calculated the intensities of weak emission lines of 
\cii, \ni, \nii, \oi\ and \oii\ using a consistent nebular model 
and SED, 
which proved successful in previous 
work in reproducing the intense emission lines and overall 
density diagnostics and surface brightness maps of \ic. 
We compared our calculations with a published deep spectroscopic 
survey taking into account the fine structure of the 
lines and including some weak lines with possible blends. 
Although this nebula shows an almost spherical symmetry, 
our calculations do take into account the position, 
size and orientation of the aperture used in the observations. 

As in our previous work on the Orion nebula, 
we found that fluorescence is responsible for the excitation of 
lines from s, p and d states in \nii. 
Fluorescence also contributes significantly to the excitation 
of most lines of the s, p and d states in \cii\ 
and some p and d states of \oii\  
while the rest of the \oii\ states are mostly excited 
by recombination. 
Some lines excited by fluorescence are sensitive diagnostics 
of the details of the SED, the expansion velocity gradient 
and turbulence of the nebula, the transition probabilities 
involved in the pumping mechanism, 
and their intensity is closely related to lines in the 
stellar atmosphere. 
The recombination rates can explain most of the 
observed intensities from f and g states of all species 
in the ionized region although there are 
many observed transitions with excited core configurations 
awaiting the calculation of their effective recombination coefficients. 

Ionic abundances deduced from dipole--allowed lines assuming 
pure recombination rates without taking into account 
fluorescence excitation of the lines can be 
overestimated by factors of up to 10 for \cii, 200 for \nii, 
2 for \oii, and 500 for \ni\ and \oi\ depending on the 
transition being considered, 
and a detailed nebular model and SED are needed to properly 
estimate ionic abundances from lines mostly excited by 
fluorescence. 

The automated procedure of line identification 
with the \emi\ code seems to be in accordance with 
our calculations because 
the largest discrepancies between predicted 
and observed line intensities tend to occur in 
lines with dubious identifications or blends as 
indicated by that code. 
A deep spectroscopic nebular survey coupled with an 
appropriate line identification procedure can prove 
useful to test the accuracy of quantum mechanical 
calculations of atomic data.

The \ni\ quartet and \oi\ triplet lines are produced by fluorescence 
in a narrow and dense interface of the neutral and ionized regions. 
Electron collisions compete with fluorescence in the 
excitation of the [\ni]\lam5200.26 and \lam5197.90 lines while 
fluorescence contributes a negligible amount to the 
excitation of the [\oi]\lam6300.30, 6363.78 and \lam5577.34 lines. 
However our calculations tend to underestimate 
systematically the observed intensities of 
the [\ni] lines by a factor of 5, 
and further work on the modeling of the PDR is 
necessary to fully understand the formation of those lines.

\section*{Acknowledgments} The authors wish to thank Peter Storey 
for providing additional atomic data, Sultana Nahar for providing 
atomic data in advance of publication, and Bob Williams for useful 
comments on the manuscript.
CM acknowledges support from grant CONACyT CB2010/153985 (M\'exico).

\label{lastpage}

\end{document}